\documentclass[12pt,epsfig]{article}

\usepackage{slashed}
\usepackage{graphicx}
\usepackage{amsmath}
\usepackage{placeins}

\newskip\humongous \humongous=0pt plus 1000pt minus 1000pt
  \newif\ifdtup

\def\frac#1#2{ {{#1} \over {#2} }}

\def\eg{\hbox{\em e.g. }}
\def\ie{\hbox{\em i.e. }}

\def\beq{\begin{equation}}
\def\eeq{\end{equation}}

\def\beqn{\begin{eqnarray}}
\def\eeqn{\end{eqnarray}}

\def\Tr{\mbox{Tr}\;}
\def\MSbar{\overline{\rm MS}}

\def\LL{{l}}

\begin{document}

\title
{                   
High-loop perturbative renormalization constants \\
for Lattice QCD (II): \\
three-loop quark currents for tree-level Symanzik 
improved gauge action and $n_f=2$ Wilson fermions}

\author
{M.~Brambilla and F.~Di~Renzo  \\
\small{Dipartimento di Fisica e Scienze della Terra, Universit\`a di Parma} \\
\small{and INFN, Gruppo Collegato di Parma} \\
\small{I-43100 Parma, Italy}
}

\maketitle

\begin{abstract}
Numerical Stochastic Perturbation Theory was able to get three- (and even four-) loop 
results for finite Lattice QCD renormalization constants. More recently, a conceptual 
and technical framework has been devised to tame finite size effects, which had been 
reported to be significant for (logarithmically) divergent renormalization constants. 
In this work we present three-loop results for fermion bilinears in the Lattice QCD 
regularization defined by tree-level Symanzik improved gauge action and $n_f=2$ 
Wilson fermions. We discuss both finite and divergent renormalization constants in 
the RI'-MOM scheme. Since renormalization conditions are defined in the chiral limit, 
our results also apply to Twisted Mass QCD, for which non-perturbative computations 
of the same quantities are available. \\
We emphasize the importance of carefully accounting for both finite lattice space and 
finite volume effects. In our opinion the latter have in general not attracted the attention 
they would deserve. 
\end{abstract}

\section{Introduction}

A few years ago the Parma group embarked in an ambitious program: computing renormalization 
constants for lattice QCD to three-loop accuracy. Since a non-perturbative computation 
of renormalization constants (RCs) has been the preferred choice for quite a long time, the 
rationale for such a program deserves a few words. The theoretical status of a perturbative 
computation of RCs is in principle firm: from a fundamental point of view, 
renormalizability is strictly speaking proved only in Perturbation Theory (PT) and quantities 
like fermion bilinears are either finite or only logarithmically
divergent; since there are no power divergences, PT must work. From a practical 
point of view difficulties show up: traditional (diagrammatic) Lattice PT is 
cumbersome, so that one can get at most two-loop results; in practice, most results are 
only one-loop. Since Lattice PT itself is badly convergent, this is a serious concern. 
Numerical Stochastic Perturbation Theory (NSPT \cite{NSPT0,NSPT1}) enables high-loop computations 
and circumvents this problem. Indeed, three- (and even four-) loop results were published 
for finite RCs in the scheme defined by Wilson action and Wilson
fermions \cite{NSPT_Zs}. \\

\noindent A follow-up of \cite{NSPT_Zs} was announced for logarithmically divergent currents, for 
which a careful assessment of finite size effects is needed. In recent years a clean way 
to effectively control the latter was introduced in \cite{NSPT_Gh,NSPT_Gl}, which we put at work 
also here. This is the first of two papers dealing with the three-loop computation of Lattice 
QCD RCs in the RI'-MOM scheme for plain Wilson fermions and improved gauge actions: in the 
present paper we report results for the tree-level Symanzik improved
gauge action with $n_f=2$ and discuss the general framework of 
finite lattice spacing and finite size corrections; 
in a second one we deal with Iwasaki gauge action with $n_f=4$
and tackle the all the problems connected with summing PT series for Lattice QCD RCs \cite{NSPT_Iwa}. 
Updates on these computations have been presented in recent years at the Lattice 
conferences, and in particular preliminary results were quoted in \cite{NSPT_Zs_LAT2012}. 
We emphasize that in both cases ($n_f=2$ tree-level Symanzik and $n_f=4$ Iwasaki) results 
can be compared with analogous non-perturbative computations for Twisted Mass fermions 
\cite{ETMC_Zs,ETMC_ZsIWA}: since the renormalization scheme is massless, RCs are the same. 
In another paper we will finally fill the gap which was left in \cite{NSPT_Zs} for 
RCs of logarithmically divergent currents for Wilson fermions and Wilson gauge action. \\

\noindent The overall structure of this paper is as follows:
\begin{itemize}
\item RI'-MOM is the scheme we adhere to. Section 2 recalls the basic definitions and 
points out a crucial issue for the success of our computations: the logarithmic 
contributions to the RCs we will compute can be got from continuum computations which 
are available in the literature. 
\item From the discussion of Section 2 it will be clear that a two-loop matching of 
continuum and lattice scheme is needed. Since this is not available in the literature 
for the gauge action at hand (tree-level Symanzik), we derive it in Section 3. We comment on 
the level of accuracy which we can attain, discuss in which sense this is enough and put 
forward the strategy for a better determination. 
\item No chiral extrapolation is needed in our computations. In PT staying at zero quark 
mass is enforced by inserting the proper counterterms: in Section 4 we present our three-loop 
result for the Wilson fermions critical mass for the tree-level Symanzik action (with $n_f=2$). 
This section also contains a few technical details on our computations
({\em}e.g. the number of configurations which were generated).
\item The extraction of the continuum limit is attained by fitting irrelevant contributions, 
which should be compliant to the lattice symmetries\footnote{The use 
of hypercubic symmetry has been widely worked out also by the Orsay 
group; see \eg \cite{French}.}. This should be done having in mind 
that RI'-MOM is defined in the infinite volume: there is a subtle interplay between 
fitting finite lattice spacing and finite volume corrections. Section 5 is devoted to a general 
discussion of our strategy to get results in the continuum and in the infinite volume limit.
\item Section 6 contains our results; in particular, we briefly
  comment on the comparison with non-perturbative results. 
\end{itemize}
As said, this paper has a follow-up in 
\cite{NSPT_Iwa}, which contains more
remarks on different ways of summing the series, trying to single out
the different (relevant and irrelevant) contributions. In
\cite{NSPT_Iwa} we also comment to which extent the techniques we
put at work in the NSPT context can provide a fresher look into the
lattice version of the RI'-MOM scheme.

\section{RI'-MOM and its logarithms}

RI'-MOM is one of the most popular renormalization schemes for Lattice QCD \cite{RI-MOMrm}; 
being regulator independent, it can be effectively adopted in a lattice regularization. While this has 
been highly recognized, one technical detail has been not yet fully
appreciated: in a RI'-MOM perturbative computation of lattice RCs,
logarithmic contributions can be inferred from continuum
computations. This is extremely useful to us. In a traditional
computation, logarithmic contributions are the {\em easy} part, while
finite terms require the really big efforts; in NSPT it is just the other
way around. As it will be clear in the following, we need to fit our
results to single out relevant and irrelevant contributions. While
disentangling logarithmic and finite terms is in principle feasible,
this would require a terrific numerical precision, {\em de facto}
impossible to attain. \\

To renormalize quark bilinears in RI'-MOM one starts from Green functions constructed 
as expectation values computed on external quark states at fixed momentum $p$ 
\[
G_{\Gamma}(p) \, = \, \int dx \,\langle p | \; \overline{\psi}(x) \Gamma \psi(x) \; | p \rangle 
\]
By inserting different $\Gamma$ one obtains the Green functions
relevant for the different currents, {\em e.g.} the scalar (identity),
pseudoscalar ($\gamma_5$),  vector ($\gamma_{\mu}$), axial
($\gamma_5\gamma_{\mu}$). 
Since these are gauge-dependent quantities, a choice for the gauge has
to be made. As it is common for the lattice implementation of RI'-MOM, Landau
gauge is our choice. This is mainly due to technical reasons, since Landau gauge can
be enforced in a lattice simulation by a minimization procedure. For
NSPT the same holds, including the additional choice of Fast Fourier
Transform acceleration (see \cite{NSPT0,NSPT1}). 
From Green functions, vertex functions are then obtained by amputation ($S(p)$ is the quark propagator)
\[
\Gamma_{\Gamma}(p) = S^{-1}(p) \, G_{\Gamma}(p) \, S^{-1}(p). 
\]
 The quark field renormalization constant has to be computed from the condition 
\begin{equation}
Z_q(\mu,\alpha) = -i\frac{1}{12}\frac{Tr(\slashed{p}S^{-1}(p))}{p^2}|_{p^2=\mu^2}.
\end{equation}
After projecting on tree-level structure 
\beq
O_\Gamma(p) = Tr\left(\hat P_{O_\Gamma}\Gamma_\Gamma(p)\right), 
\eeq
one enforces renormalization conditions that read 
\begin{equation}\label{master}
Z_{O_\Gamma}(\mu,\alpha) Z_q^{-1}(\mu,\alpha)O_\Gamma(p)|_{p^2=\mu^2} = 1.
\end{equation}
In order to have a mass-independent scheme, all this is defined at zero quark mass.

\noindent In a lattice perturbative computation, we can write the generic $Z$ in the continuum limit ($a \rightarrow 0$)
\begin{equation}
Z (\mu,\alpha) = 1+\sum_{n>0}d_n(\LL)\,\alpha(\mu)^n\qquad d_n(\LL)=\sum_{i=0}^nd_n^{(i)}\LL^i \qquad \LL\equiv\log(\mu a)^2
\end{equation}
where the lattice cutoff ($a$) is in place and the expansion is in the renormalized coupling. 
To make our notations a bit lighter, we have omitted any reference to any
operator, {\em i.e.} we wrote $Z$ and not
$Z_{O_\Gamma}$ as we did a few lines above. In the same spirit we have
omitted any dependence on the (covariant) gauge parameter $\lambda$:
as already pointed out, the reader has to assume the choice $\lambda=0$ (Landau). 
Making this choice explicit simplifies the formulas; we
will comment on the general formulation a few lines below, 
pointing out why our (apparently naive) notation is indeed correct for
Landau gauge.
For finite quantities ({\em e.g.} vector and axial 
currents) $d_n^{(i>0)}=0$; for divergent quantities ({\em e.g.} scalar and pseudoscalar 
currents) divergencies show up as powers of $\log(\mu a)^2$. To compute the $Z$s in NSPT, we
want to eventually manage expansions in the bare lattice coupling $\alpha_0$
\begin{equation}
Z (\mu,\alpha_0) = 1+\sum_{n>0}\overline{d}_n(\LL)\,\alpha_0^n \qquad \overline{d}_n(\LL)=\sum_{i=0}^n\overline{d}_n^{(i)}\LL^i.
\end{equation}

By differentiating Eq.~(4) with respect to $\log(\mu a)^2$ 
one obtains the expression for the anomalous dimension 
\[
\gamma = \frac{1}{2} \frac{d}{d\LL} \log Z.
\]
Since this is a scheme dependent, finite quantity, one has to recover
an expansion in which coefficients are finite numbers (with no 
dependence on the regulator left)
\begin{equation}
\gamma = \sum_{n>0}\gamma_n \,\, \alpha(\mu)^n.
\end{equation}
This expansion is known to three-loop accuracy from continuum
computations \cite{Gracey}. 
Imposing that the expression obtained by differentiating Eq.~(4)
matches Eq.~(6) (with the proper values for the $\gamma_n$ read from \cite{Gracey}) 
we can obtain the expressions of all the $d_n^{(i >0)} \, (n\leq 3)$. In practice the
solution comes from the request that all the (powers of the)
logarithms cancel out; as a result, each $d_n^{(i>0)}$ is expressed in terms
of the $\gamma_{m\leq n}$, the $d_{m \leq n}^{(0)}$\footnote{This
  dependence is not a problem, since we solve for any quantity order
  by order ({\em i.e.} everything at order $m<n$ is known when one
  determines a quantity at order $n$). Notice 
that a similar dependence holds for the $\overline{d}_n^{(i>0)}$ ; they
depend on $\overline{d}_{m \leq n}^{(0)}$. 
As it will be clear in Section 5, once a $\overline{d}_k^{(0)}$ has been
determined, its value can enter the analysis at higher orders.} 
and the coefficients of the $\beta$-function; the latter come into place since part of
the dependence on $\mu$ in Eq.~(4) is via the coupling $\alpha(\mu)$.\\

The $\overline{d}_n^{(i>0)}$ of Eq.~(5) can finally be obtained by re-expressing the
expansion in Eq.~(4) as an expansion in the bare 
coupling $\alpha_0$. This makes the $\overline{d}_n^{(i)}$ depend (also) on 
the coefficients of the matching of the continuum and the lattice
couplings. In particular, our three-loop expansion of the Zs
asks for a two-loop matching of the continuum coupling to the lattice 
coupling in the scheme we are working in, {\em i.e.} tree-level
Symanzik gauge action with $n_f=2$ Wilson fermions. Since this is not known from
the literature (\cite{Aoki} provides a one-loop matching), we derived
it: this is discussed in the next section. \\

We now come back to the issue of (covariant) gauge parameter
dependence. In a generic (covariant) gauge, not only the dependence on
$\lambda$ enters Eq.~(4), but also the gauge parameter anomalous
dimension comes into place in linking Eq.~(4) to Eq.~(6). We have
worked out the formulas with generic $\lambda$ and checked the
correctness of our results with the two-loop computations of
\cite{Haris2L}, which are obtained in Feynman gauge. Notice that the
apparently naive recipe of neglecting the $\lambda$-dependence from
the very beginning (as we did in our previous discussion) returns
correct results, {\em i.e.} one obtains the same results by keeping
track of all the $\lambda$-dependences and by finally putting 
$\lambda=0$. This is due to the fact that the non trivial dependence
on the gauge parameter anomalous dimension is itself proportional to
$\lambda$. \\

The expressions for the $\overline{d}_n^{(i>0)}$ (and $d_n^{(i>0)}$) are available
upon request in the form of {\em Mathematica} notebooks; in the
following we will focus on the finite $\overline{d}_n^{(0)} \, (n\leq 3)$ and 
adhere to the standard recipe of summing the series at $\mu a =
1$. Actually we will report the results as expansions in yet another coupling, namely
$\beta^{-1} \equiv \frac{2 \pi \alpha_0}{3}$ (more on this later). 
As we have already pointed out, in order to reconstruct the whole set 
$\{\overline{d}_n^{(i>0)}\}$, the only piece of information which is missing in the
literature is the two-loop matching of continuum to lattice coupling
for the regularization at hand: we report this result in the next section.

\section{Two-loop matching of continuum and lattice couplings}

In the following we provide the matching of couplings enabling us to go from
Eq.~(4) to Eq.~(5). We will make use of the notation 
$\alpha_0 = \alpha_{TLS}$ to enlighten that the lattice coupling we are
referring to is the one defined by the regularization at hand, with
the Tree-Level Symanzik (TLS) gauge action in place\footnote{The
  choice for $n_f=2$ Wilson fermions will also be (implicitly, as for
  notation) assumed.}. The matching will be to the $\overline{MS}$
scheme, $\alpha_{\overline{MS}}$ being the coupling in which the expansions in \cite{Gracey}
are expressed (strictly speaking, it
suffices that this holds at the finite order one is interested in). \\

The general form of the matching between two schemes (unprimed and primed) reads
\begin{equation}\label{matching}
\alpha(s\mu) = \alpha'(\mu) +c_1(s) \, \alpha'(\mu)^2 +c_2(s) \, \alpha'(\mu)^3+\ldots.
\end{equation}
The coefficient $s$ accounts for the 
choice of different momentum scales; it enters the expressions for the
matching coefficients $c_1(s)$ and $c_2(s)$
\begin{eqnarray}
c_1(s) &=& 2b_0 \log\frac{\Lambda}{\Lambda'}-2b_0\log{s}\label{c1}\\ 
c_2(s) &=& c_1(s)^2-2b_1\log{s}+2b_1
\log\frac{\Lambda}{\Lambda'}+\frac{b_2-b_2'}{b_0}.\label{c2}
\end{eqnarray}
Here $b_0, b_1, b_2$ and $b_2'$ are coefficients of the $\beta$-function, 
while $\Lambda$ and $\Lambda'$ are the scales associated with the
two regularizations. While $b_0$ and $b_1$ are
universal, $\Lambda$ and $b_2$ depend on the scheme (and so they come
in both primed and unprimed versions\footnote{One could object the 
notation is a bit sloppy: in our notation $b_2$ is unprimed as referring to the
unprimed scheme and NOT because it is universal.}). 
Notice that Eq.~(\ref{c2}) states that the two loop matching of $\alpha_{TLS}$ to 
$\alpha_{\overline{MS}}$ also entails the knowledge of $b_2^{TLS}$, 
since $b_2^{\overline{MS}}$ is known.

\subsection{Strategy for the matching}
There is no obvious way of computing the direct matching of $\overline{MS}$ to
the TLS scheme making use of NSPT. We will go through the
strategy of first matching to an intermediate scheme. Once again, we
will rely on the fact that no computation of logarithms will be
needed: every relevant logarithmic dependence is known. 
The strategy has already been used in~\cite{residual_massQ, residual_massU}
(although in those works we had another goal). \\

In the lattice regularization defined by TLS gauge action and $n_f=2$
Wilson fermions, we computed the perturbative expansions of
rectangular Wilson loops $W(R,T)$  (of extensions $R$ and $T$) and 
from those we computed logarithms of Creutz ratios  
\[
V_T(R) = \log\left(\frac{W(R,T-1)}{W(T,R)}\right).
\]
Notice that in the previous formula everything is dimensionless. In
particular, $R$ and $T$ are measured in lattice units ({\em i.e.} they
are integer numbers): at a given, fixed value of the lattice spacing $a$, 
physical lengths associated to them are $r=Ra, \, t=Ta$.
The static quark potential can be defined via
\[
a V(r) = a V(R a) = \lim_{T\to\infty}V_T(R).
\]

The static quark potential is the quantity which describes the
interaction energy of a infinitely heavy $q\bar q$ pair at a distance $r$, which 
in its full (non-perturbative) form is in first approximation just 
the sum of a string tension contribution, 
which is responsible for confinement, and a $r^{-1}$ contribution, 
whose interpretation is different in different IR/UV regimes 
\[
V(r) = \frac{C}{r}+\sigma r.
\]
In PT the first term is just the Coulomb potential and there is no
string tension. In a lattice regularization, one is
left in addition with a linearly divergent term, which gives the
so called residual mass of the heavy quark:
\begin{equation}\label{ValphaR}
a V(r) = a V(R a) = 2\delta m-C_F\frac{\alpha_V(r^{-1})}{R}.
\end{equation}
While $\delta m$ is associated to a linearly divergent quantity, logarithmic divergencies 
are absorbed in $\alpha_V$; extra ({\em corners}) divergences are absent
because the quantity is built out of ratio of rectangular loops. 
Eq.~(\ref{ValphaR}) defines the potential coupling $\alpha_V(r^{-1})$ we will be 
concerned with. Notice that the perturbative computation of the 
(power divergent) residual mass is not supposed to be a reliable one. \\
  
A perturbative computation of the static quark potential in our 
lattice scheme reads
\begin{eqnarray}\label{ValphaLat}
a V(r) = a V(R a) & = &V_0(R) \, \alpha_{TLS} + V_1(R) \, \alpha_{TLS}^2 + V_2(R) \,
\alpha_{TLS}^3 + \mathcal{O}(\alpha_{TLS}^4)\\ \nonumber
& = & 2 \left( \delta m_0 \, \alpha_{TLS} + \delta m_1 \,
  \alpha_{TLS}^2 + \delta m_2 \, \alpha_{TLS}^3 + \mathcal{O}(\alpha_{TLS}^4) \right) + \\ \nonumber
& & -C_F\frac{\alpha_{TLS}}{R}\left(1 + C_1(R)\, \alpha_{TLS}+
C_2(R)\, \alpha_{TLS}^2 + \mathcal{O}(\alpha_{TLS}^3)\right),
\end{eqnarray}
where subscripts are written according to the actual loop counting. 

In order to trade the description in terms of the $V_i(R)$ for that in
terms of $\delta m_i$ and $C_i(R)$ one has to disentangle constant and
$R^{-1}$ contributions: in NSPT this requires a fitting procedure. 
The description in terms of $\delta m_i$ and $C_i(R)$ is the
order by order version of  Eq.~(\ref{ValphaR}): we have actually
computed $\alpha_V(r^{-1})$ as an expansion 
\begin{equation}\label{matchingAVA0}
\alpha_V(r^{-1}) = \alpha_{TLS} + C_1(R)\, \alpha_{TLS}^2+ C_2(R)\, \alpha_{TLS}^3 +
\mathcal{O}(\alpha_{TLS}^4).
\end{equation}

To be more precise, this is simply a
particular case of the matching~(\ref{matching}): it is the matching
of the continuum coupling $\alpha_V(r^{-1})$ to the lattice coupling
$\alpha_{TLS}$. Since the latter is defined at the scale $a^{-1}$
while $\alpha_V(r^{-1})$ is defined at the scale $r^{-1}=a^{-1} R^{-1}$, 
the factor $s$ of Eq.~(\ref{matching}) here reads $s=R^{-1}$. In other
words, we can read the $C_i(R)$ from Eq.~(\ref{c1}) and Eq.~(\ref{c2}) 
({\em i.e.} $C_i(R) = c_i(R^{-1})$)
\begin{eqnarray}
C_1(R) & = & 2b_0 \log\frac{\Lambda_V}{\Lambda_{TLS}}+2b_0\log{R}\label{c1V0}\\
C_2(R) &=& C_1(R)^2+2b_1\log{R}+2b_1
\log\frac{\Lambda_V}{\Lambda_{TLS}}+\frac{b_2^{(V)}-b_2^{(TLS)}}{b_0}.\label{c2V0}
\end{eqnarray}

Notice that everything has been written in the limit $a \rightarrow 0$
(and in the infinite volume limit, as it is clear from the definition
of $aV(Ra)$ in terms of $V_T(R)$). We will have to come back to this 
when we discuss the NSPT computation. \\

As a byproduct of our computation we also obtain the residual mass
$\delta m$ as an expansion in $\alpha_{TLS}$ 
\begin{equation}\label{deltaM}
\sum_{n\ge0}\delta m_n \, \alpha_{TLS}^{n+1}.
\end{equation}

Once we have computed the matching between $\alpha_V(R)$ and $\alpha_{TLS}$, we
need the matching of $\alpha_V(R)$ to $\alpha_{\overline{MS}}$. This can be 
read from the computation in~\cite{York}. 

\subsection{Results}
The NSPT computations were performed on a $32^4$ lattice; we computed the
Wilson loops $W(R,T)$ for all the values of $R$ and $T$ up to 16. 
The quark mass was set to zero by plugging 
the appropriate mass counterterm, {\em i.e.} the perturbative critical mass
as read from \cite{HarisMC} (see next Section).
Results were averaged over $\sim$150 lattice
configurations\footnote{The reader will notice that we took more WL
  measurements than current measurements on $32^4$.}. 

From the $W(R,T)$ we computed the $V_T(R)$. We could not take the
$T\to\infty$ limit, but regarded the $V_T(R)$ as our estimate for $aV(Ra)$. 
This is a first (finite volume) approximation in our setting. By fitting 
(order-by-order) our $V_T(R)$ data to $aV(Ra)$ as defined by 
eq.~(\ref{ValphaLat}) (which is valid in the $a \rightarrow 0$ limit) 
we incurred yet another source of approximation, since no attempt was 
made to take into account irrelevant, finite $a$ effects. 
Despite the distortions expected from these lattice artifacts, we 
extracted from our data both the residual quark mass $\delta m_i$ and the expected 
$C_i(R)$ ({\em i.e.} we fitted the parameters entering their
expressions). Since most of the parameters are known, we can estimate
{\em a posteriori} how good (or bad) the procedure is.

In order to at least minimize the irrelevant effects we 
considered intervals of $R$ such that 
\begin{itemize}
\item
$T > R$ ($T/R\sim 2.5$); 
\item
$R$ itself is not too small ($R\ge 3$);
\item 
the fitting intervals themselves are from 3 up to 7 points long. 
\end{itemize}

On top of the systematic (lattice artifacts) errors, results 
are of course also affected by statistical errors. The
relative weight of these effects is different for different orders. 
This actually opens the way to a possible (careful) tradeoff between the errors.
We adopted the following strategy: when systematic effects are clearly
distinguishable (i.e. statistical effects are relatively small), we only considered
$T=16$ data (this is the case of the
tree-level potential). 
When statistical errors are significant on their own, 
the systematic (finite $T$) effect is not that clear. 
In this case we decided to neglect this systematic effect and tame the statistical 
noise by averaging over different values of $T$ (from $T = 14$ to $T = 16$). 
In this way we obtained smoother curves. Further details can be found
in \cite{alphaMATCHlat2012}. 

In order to verify the reliability of our approach we first checked
known results. Notice that despite the coupling parameters are known, 
even lower order results are not trivial, since at any
order residual mass is unknown (we got it as a byproduct). 
In Figure~1 we show our estimates for tree-level, one-loop and two-loop 
potential. We estimated $3\le R\le 7$ as the best fitting interval for tree-level
and one loop; the same interval was also taken for two-loop
computation.

Tree-level data were fitted to the functional form
\[
V_0(R) = 2\delta m_0 - \frac{C_F}{R}
\]
obtaining $\delta m_0 = 1.84\pm 0.01$, while $C_F$ is reconstructed 
to a few percent. This gives a first rough idea 
of the impact of systematic effects. \\

\begin{figure}[!tb]
  \hspace{-0.7cm}
  \begin{tabular}{ccc}
     \includegraphics[height=4.3cm,clip=true]{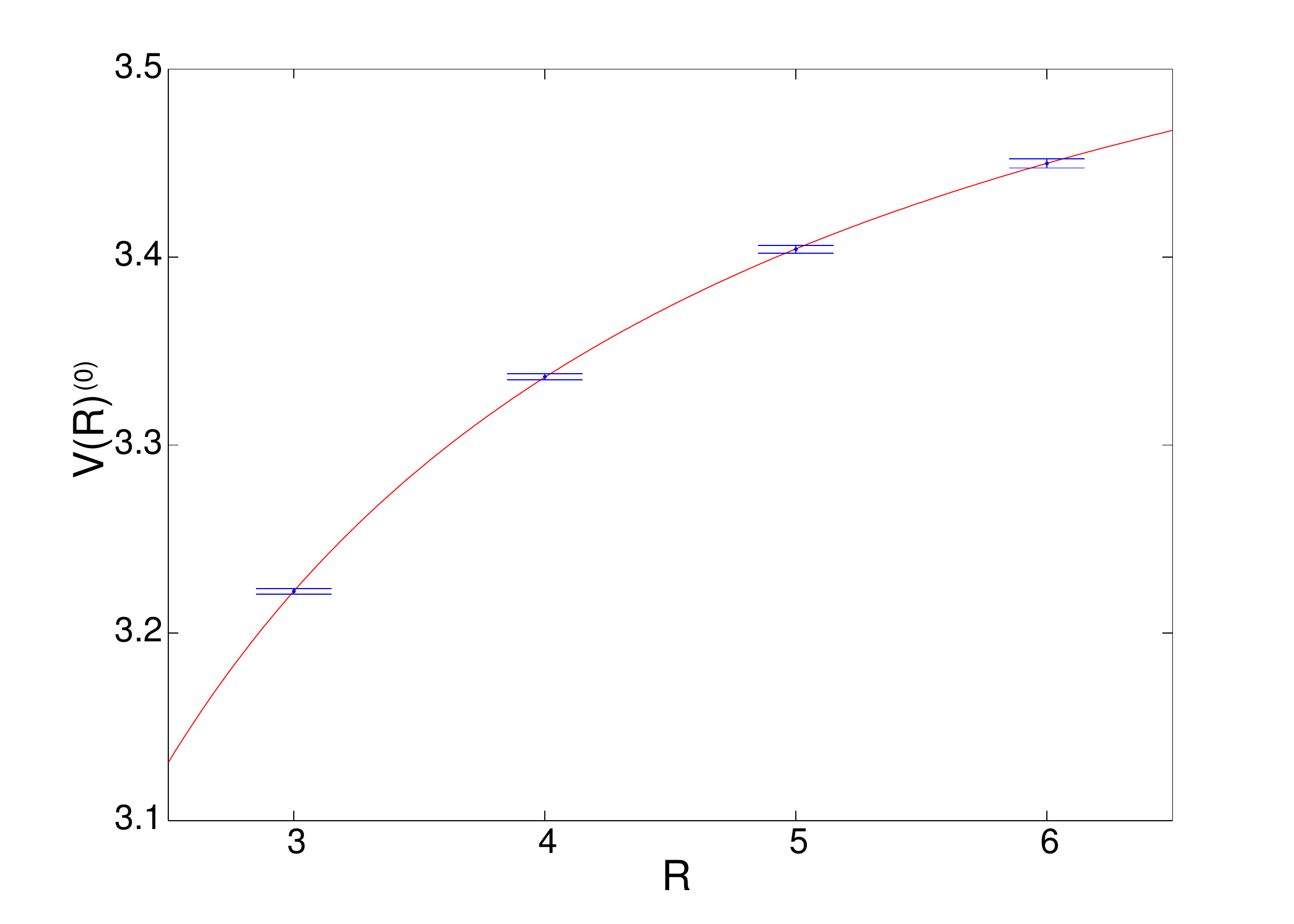}
     &
  \hspace{-0.9cm}
     \includegraphics[height=4.3cm,clip=true]{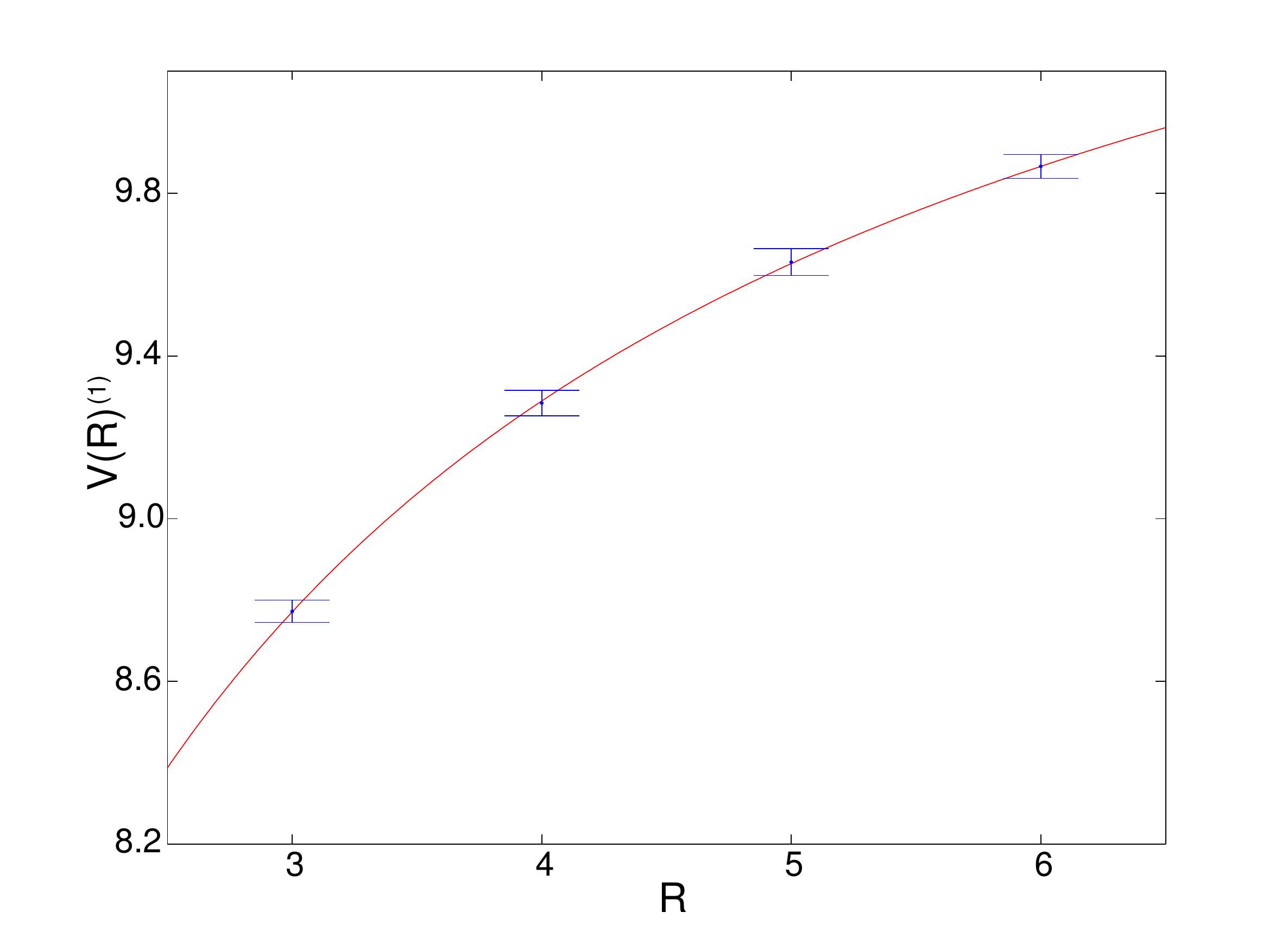}
     &
  \hspace{-0.9cm}
     \includegraphics[height=4.3cm,clip=true]{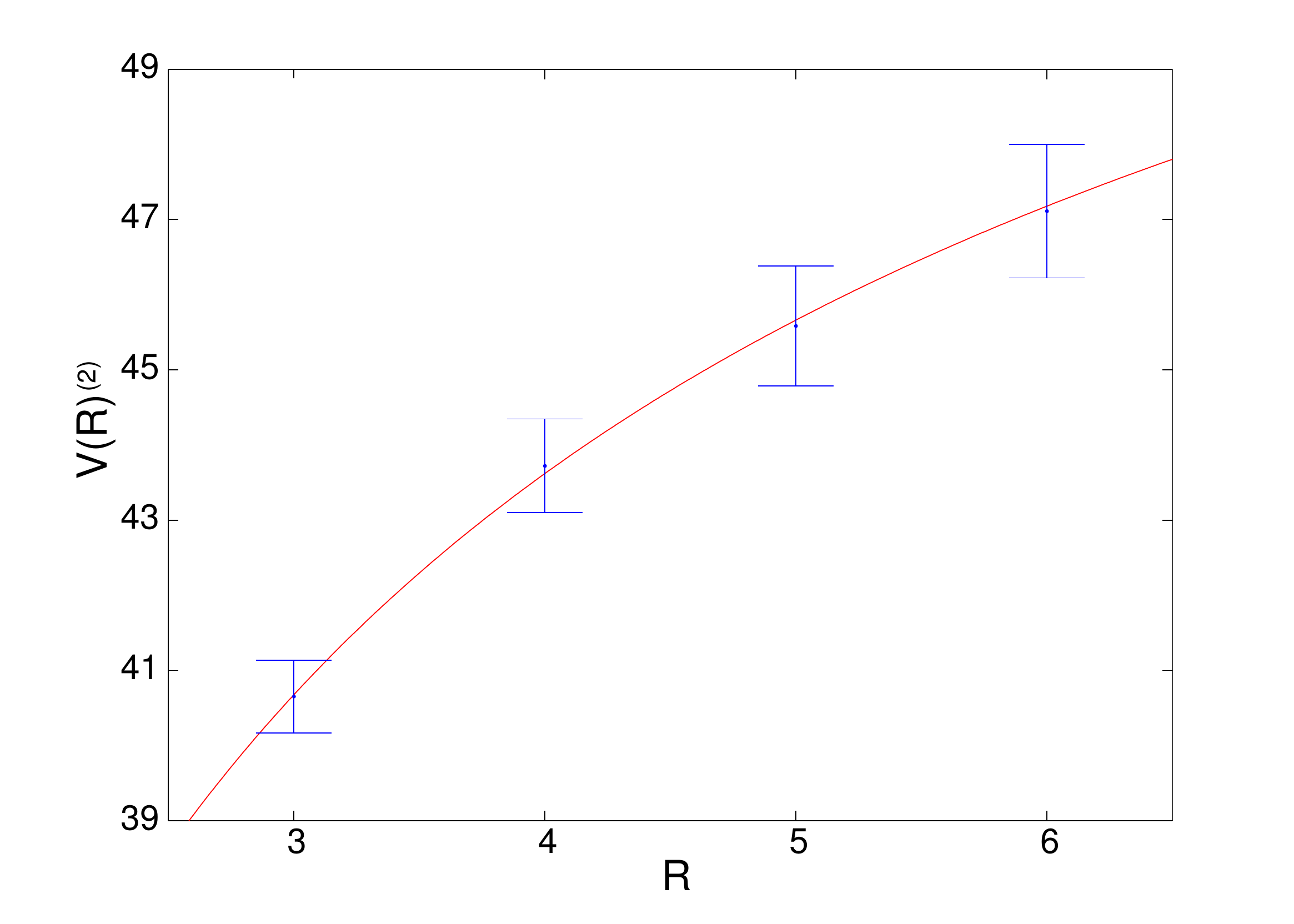}
  \end{tabular}
  \vspace{-2mm}
  \caption{Data and fit (continuous line) for (from left to right)
    tree-level, one-loop, two-loop potential.}
\end{figure}

\noindent At one-loop, we plugged the analytical value for $C_F$ and
tried to fit the constant term $\log\frac{\Lambda_V}{\Lambda_{(TLS)}}$ 
in the functional form
\[
V_1(R) = \delta m_1 - \frac{C_F}{R}2b_0\left(\log
R+\log\frac{\Lambda_V}{\Lambda_{TLS}} \right).
\]
We obtained $\log\frac{\Lambda_V}{\Lambda_{(TLS)}} = 2.8\pm 0.1$, to be
compared to the analytical result $2.8191$~\cite{Aoki}. We 
also obtained $\delta m_1 = 5.71 \pm 0.01$. 

\noindent At two-loop we finally tackled the determination of the
quantity we are interested in; the functional form 
\[
V_2(R) = \delta m_2 - \frac{C_F}{R}\left( c_1(R)^2 + 2b_1\log
R + 2b_1\log\frac{\Lambda_V}{\Lambda_{TLS}} +
\frac{b_2^{(V)}-b_2^{(TLS)}}{b_0} \right).
\]
depends on the unknown $\delta m_2$ and $\frac{b_2^{(V)}-b_2^{(TLS)}}{b_0}$. 
As one can see from the figure, at two-loop fluctuations are larger than
at lower orders, and as a consequence the fit suffers from a larger indetermination. In
this case we obtained $\delta m_2 = 30\pm 1$ and
\[
  \frac{b_2^{(V)}-b_2^{(TLS)}}{b_0} \equiv X = 4 \pm 1
\]
where we have introduced a notation ($X$) we will make use of later.
Though the relative error in this value is high, we must emphasize 
that one is interested in the final matching to
$\alpha_{\overline{MS}}$. From \cite{York} we can get the matching of 
$\alpha_V$ to $\alpha_{\overline{MS}}$, and the final
result is (remember that this holds for $n_f=2$, with Wilson
regularization for the lattice fermions)
\[
\alpha_{\overline{MS}} = \alpha_{TLS} + 2.79866 \, \alpha_{TLS}^2 +
(11.5 \pm 1.0) \, \alpha_{TLS}^3 + \mathcal{O}(\alpha_{TLS}^4),
\]
where the relative error on the second coefficient is slightly less than 10\%. 

Notice that in order to assess the effectiveness of our computation, 
this is not yet the end of the story. What we are really interested in
is how the parameter $X$ (which fixes this matching) enters the 
coefficient $\overline{d}_3^{(1)}$ of Eq.~(5). Actually in the following we will 
report our results as expansions in the lattice coupling $\beta^{-1}$
(we specify the definition to the case at hand, \ie SU(3))
\[
\beta^{-1} \equiv \frac{g_0^2}{6} \equiv \frac{2 \pi \alpha_0}{3}
\]

\begin{equation}
Z (\mu,\beta^{-1}) = 1+\sum_{n>0}z_n(\LL)\,\beta^{-n} \qquad
z_n(\LL)=\sum_{i=0}^n z_n^{(i)}\LL^i  \qquad \LL\equiv\log(\mu a)^2
\end{equation}
We now show how the parameter $X$ enters $z_{S\,3}^{(1)}$,
where the extra subscript $S$ indicates that we are taking the example
of the renormalization of the scalar current: 
\[
z_{S\,3}^{(1)}  = 1.7823 + 0.0693\, X + 0.7366 \, z_{S\,1}^{(0)} + 0.3040 \, z_{S\,2}^{(0)}
\]
This is the coefficient in front of the simple log in the
three-loop order of the renormalization constant of the scalar
current; a similar relation is in place for the pseudoscalar current. 
Apart from $X$, the only parameters in the formula are 
$z_{n<3}^{(0)}$ (all other numerics have been worked out explicitly): $z_{S\,1}^{(0)} =
-0.6893$ is known analytically (\cite{Aoki}), while at two-loop we got 
$z_{S\,2}^{(0)} = -0.777(24)$\footnote{This is an example of
  what we have already pointed out: a two-loop $z_2^{(0)}$ enters the
  expression for a $z_3^{(1)}$, but since two-loop $z_2^{(0)}$ can be computed
  before we are concerned with the $z_3^{(i)}$, this is not a problem.}. 
We can conclude the indetermination which is carried by $X$
is acceptable (namely, less than $10\%$). For the pseudoscalar current
numerics is less favorable\footnote{This is due to the values of one-
  and two-loop constants $z_{P\,1}^{(0)}$ and $z_{P\,2}^{(0)}$; all
  this is of course merely a numerical accident.}, but the
indetermination remains acceptable (namely, of order $10\%$). 

One could wonder whether NSPT can do better that what we showed in
computing lattice to continuum coupling matchings. The answer is yes, a natural 
candidate for a more effective intermediate scheme being a finite 
volume one, {\em e.g.} the SF (Schroedinger
Functional) scheme. A robust NSPT formulation of PT for the SF has
been set up recently \cite{NSPT_SF}.

\section{The three-loop critical mass}

For Wilson fermions staying at zero quark mass amounts to the
knowledge of the critical mass. In perturbation theory 
the latter has to be computed at the convenient
order and plugged in as a counterterm: one does not need to go 
through an extrapolation process to reach the chiral limit (which can be a heavy task in a
non-perturbative computation, in the end always acting as a source
of systematic error). 

In order to compute three-loop renormalization constants the effect of 
the critical mass has to be corrected up to two-loop
order. Though it is not relevant for the computation at hand, 
we get as a by-product the value of the three-loop
critical mass (which is a new result). 

\begin{figure}[!htb]
\begin{center}
  \begin{tabular}{cc} 
     \includegraphics[height=6.5cm,clip=true]{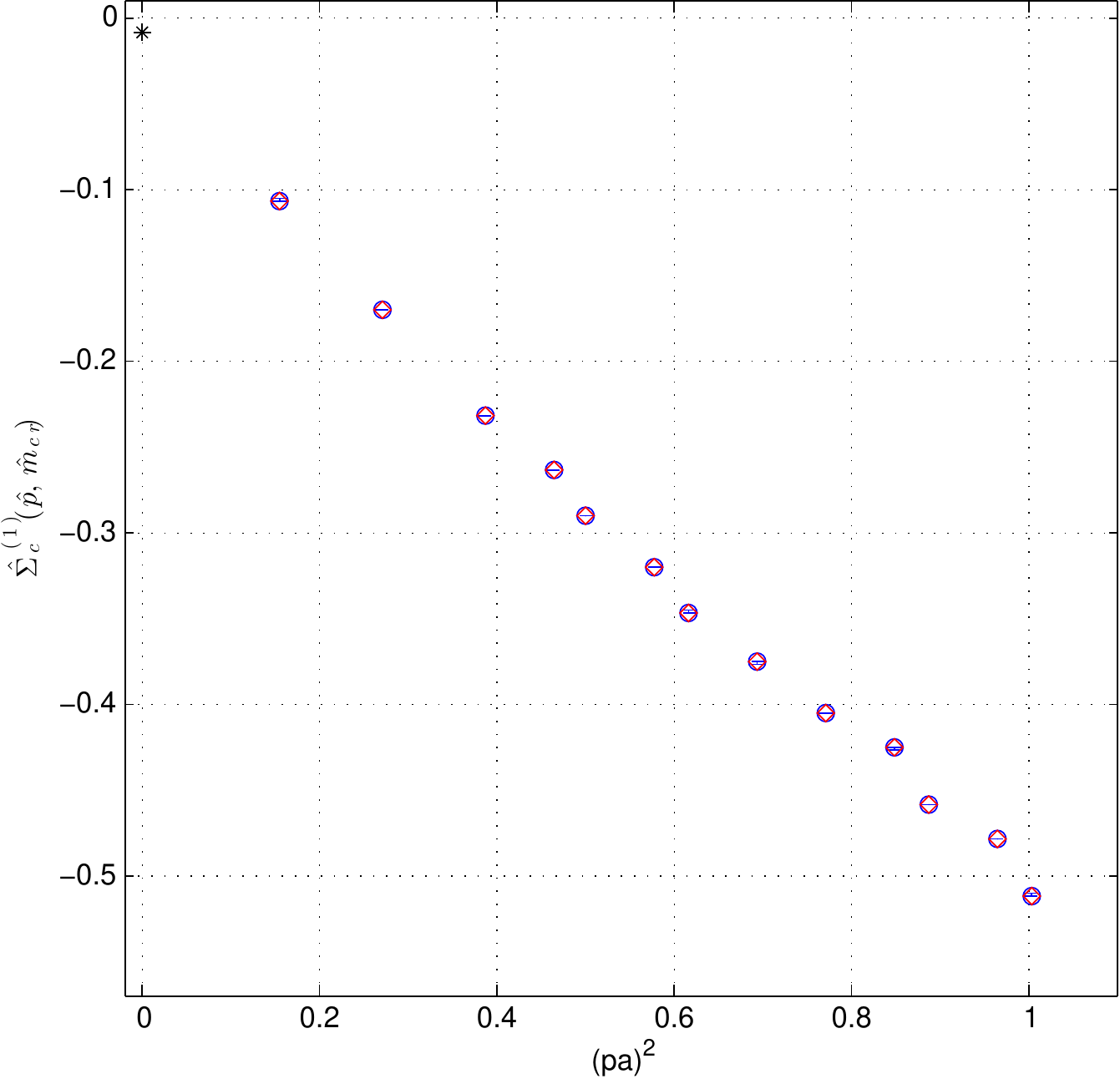}
     &
 \hspace{1.cm}
     \includegraphics[height=6.5cm,clip=true]{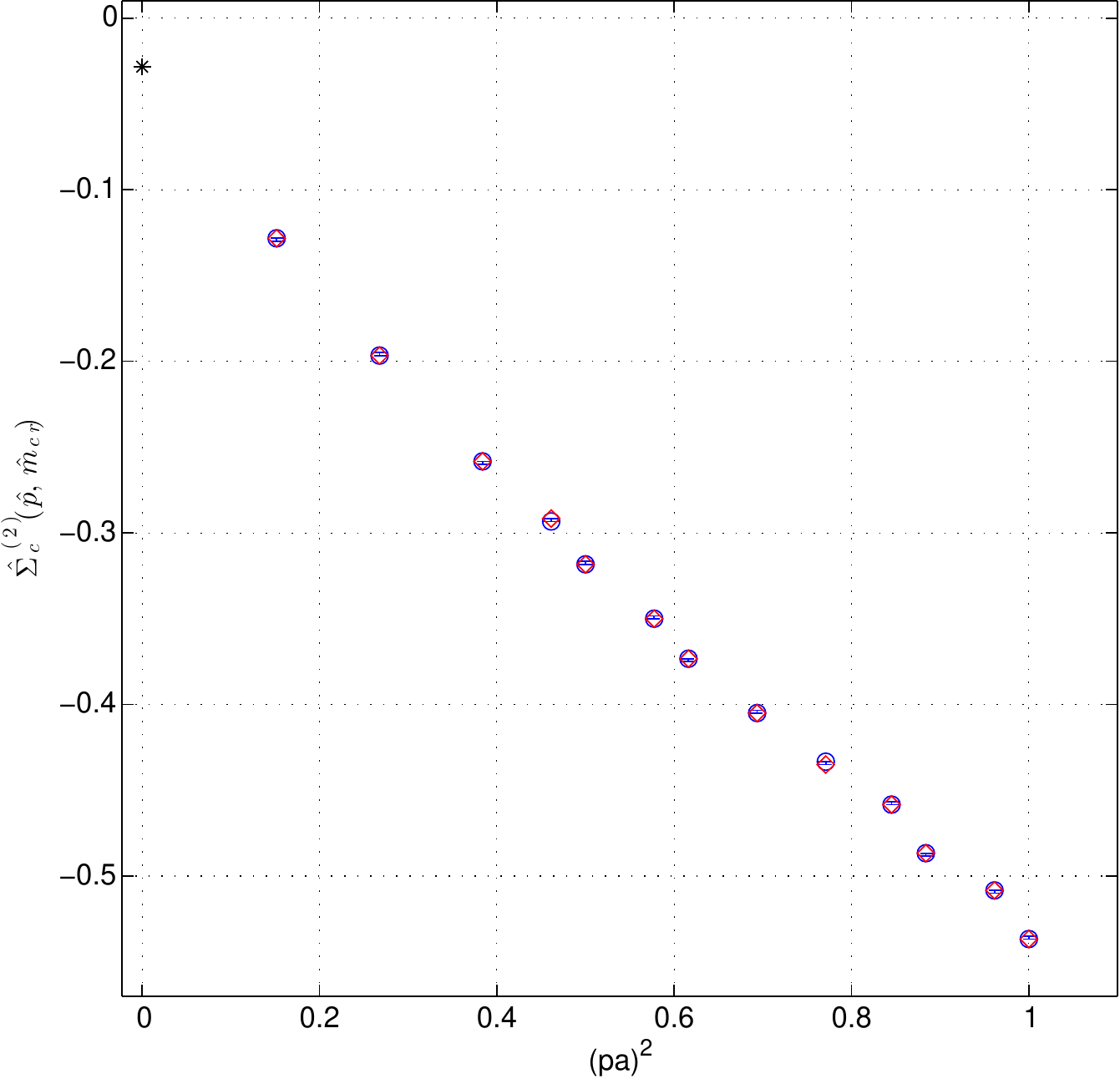}
  \end{tabular}
\end{center}
  \caption{One-(left) and two-(right)loop
    $\hat{\Sigma_c}(\hat{p},\hat{m}_{cr},\beta^{-1})$ on a $32^4$
    lattice: the (extrapolated) one- and two-loop values at
    $\hat{p}=0$ should be zero, because of counterterms, but 
they are not, because of finite volume.}
\end{figure}

\FloatBarrier

The critical mass is computed from the inverse quark propagator 
\beq
a\Gamma_2(\hat p, \hat m_{cr}, \beta^{-1}) = aS(\hat p, \hat m_{cr}, \beta^{-1})^{-1} 
= i\hat{\slashed{p}}+\hat m_{W}(\hat p) - \hat\Sigma(\hat p, \hat m_{cr}, \beta^{-1}).
\eeq
In order to have less factors $a$ in place, we have here introduced
a hat notation to denote dimensionless quantities ({\em e.g.} $\hat p
= pa$): explicit factor of $a$ will be later singled out if
needed. As already pointed out, $\beta^{-1}$ is 
the expansion parameter we adopt (and hence the dependence we quote). 
$\hat{m}_W(\hat{p}) = {\cal O}(\hat{p}^2)$ is the (irrelevant) mass term generated at 
tree level by the Wilson prescription. 

The dimensionless self-energy $\hat{\Sigma}(\hat{p},\hat{m}_{cr},\beta^{-1})$  
(which is ${\cal O}(\beta^{-1})$) reads 
\beq\label{sigma}
\hat\Sigma(\hat p, \hat m_{cr}, \beta^{-1}) = \hat\Sigma_c(\hat p,
\hat m_{cr}, \beta^{-1})+\hat\Sigma_{\gamma}(\hat p, \hat m_{cr},
\beta^{-1})+
\hat{\Sigma}_{\rm other} (\hat p, \hat m_{cr}, \beta^{-1}),
\eeq
where we have singled out the component along the (Dirac space)
identity
\[
\hat\Sigma_c(\hat p, \hat m_{cr}) = 1/4 {\rm Tr_{spin}}(\hat{\Sigma}), 
\]
and the one along the gamma matrices
\[
\frac{1}{4} \sum_{\mu} \gamma_{\mu} {\rm Tr_{spin}}( \gamma_{\mu} \hat{\Sigma} ) = \hat{\Sigma}_{\gamma},
\]
while $\hat{\Sigma}_{\rm other}$ includes all other possible contributions along the remaining elements of
the Dirac basis: these quantities are irrelevant and are always
neglected in our analysis. The critical mass can be read from
$\hat{\Sigma}_c$ at zero momentum 
\begin{equation}\label{Mc}
	\hat{\Sigma}(0,\hat{m}_{cr},\beta^{-1}) = \hat{\Sigma}_c(0,\hat{m}_{cr},\beta^{-1}) = \, \hat{m}_{cr}.
\end{equation}
Notice that restoring physical dimensions one recognizes that the
critical mass is order $a^{-1}$, and so it must be cured by an additive
counterterm. 
Since 1-loop and 2-loop orders are known \cite{HarisMC}, we plug their values as 
counterterms in our computations. As a result, a plot of one- and two-loop
$\hat{\Sigma}_c$ vs momentum should display a zero intercept in zero. 
Actually this is not strictly speaking correct; on finite lattices one
inspect corrections (which get smaller and smaller as the lattice size 
increases): see Fig.~2 for $32^4$ measurements. 

The $p=0$ intercept of
$\hat{\Sigma_c}(\hat{p},\hat{m}_{cr},\beta^{-1})$) comes from a 
fitting procedure: this is a general
feature of our computations, as it will be clear in the next
section. Since we want to remove the finite size effects, we 
perform our computations on different lattice sizes. 

\subsection{Data sets}

Measurements were taken on different sizes: $32^4$,  $20^4$,  $16^4$,
$12^4$. NSPT prescribes the numerical (order by order) integration of
the Langevin equation (see \cite{NSPT0, NSPT1}). In this work we adopt 
the simplest numerical scheme, {\em i.e.} Euler scheme. 
To remove the effects of the finite time step $\epsilon$
an extrapolation $\epsilon \rightarrow 0$ 
(in this scheme a linear one) is needed. 
In view of this, measurements were taken on configurations generated at different values of 
$\epsilon$. Table~1 summarizes our statistics. The
procedure is pretty the same as for non-perturbative simulations:
configurations are generated on which one can later measure different
observables. A preliminary analysis of 
the autocorrelations in place guided our choice for the frequency at
which we save configurations. Residual autocorrelation effects are of course
later accounted for in the analysis of different observables. \\
One should keep in mind that lattice sizes are actually pure numbers
$N=L/a$: the coupling is in NSPT an expansion parameter and there is no
physical value (say, in fermi) of the lattice spacing (and henceforth
of the lattice size).

\begin{table}[t]
\caption{Number of measurements at different 
value of the time step for the different lattice sizes.}
\begin{center}
\begin{tabular}{|c|c|c|c|}
\hline
lattice size $N = L/a$  & $\epsilon = 0.005$ & $\epsilon = 0.010$ & $\epsilon = 0.015$ \\
\hline
\hline
$12$ & 118                       & 115 & 119 \\
$16$ & 195                       & 136 & 184 \\
$20$ & 20                    & 31 & 41 \\
$32$ & 22                    & 19 & 22 \\
\hline
\hline
\end{tabular}\\
\end{center}
\end{table}

\subsection{Results}

For each value of the lattice size and at each order in the coupling, 
we get the zero momentum value of 
$\hat{\Sigma_c}(\hat{p},\hat{m}_{cr},\beta^{-1})$) by fitting the
latter as an expansion in 
hypercubic invariants, \eg
\begin{equation}
\sum_\nu \hat p_\nu^2 \;\;\;\;\;\;\;
\frac{\sum_\nu \hat p_\nu^4}{\sum_\nu \hat  p_\nu^2} \;\;\;\;\;\;\;
{(\sum_\nu \hat p_\nu^2)}^2 \;\;\;\;\;\;\;
\sum_\nu \hat p_\nu^4 \;\;\;\;\;\;\;
\frac{\sum_\nu \hat p_\nu^6}{\sum_\nu \hat p_\nu^2} \;\;\;\;\;\;\;
\ldots
\end{equation}

Figure~2 gives an idea of the effectiveness of such fits. Once we have
the $\hat{\Sigma_c}(0,\hat{m}_{cr},\beta^{-1})$ for each lattice size,
we have to extrapolate them to the infinite volume limit: 
Figure~3 displays the behavior of (one- and two-loop) results as inverse powers of 
$N=L/a$. Results are fully consistent with the (known) critical mass 
analytical values we have plugged in. \\

\begin{figure}[!tb]
\begin{center}
  \begin{tabular}{cc}
     \includegraphics[height=6.5cm,clip=true]{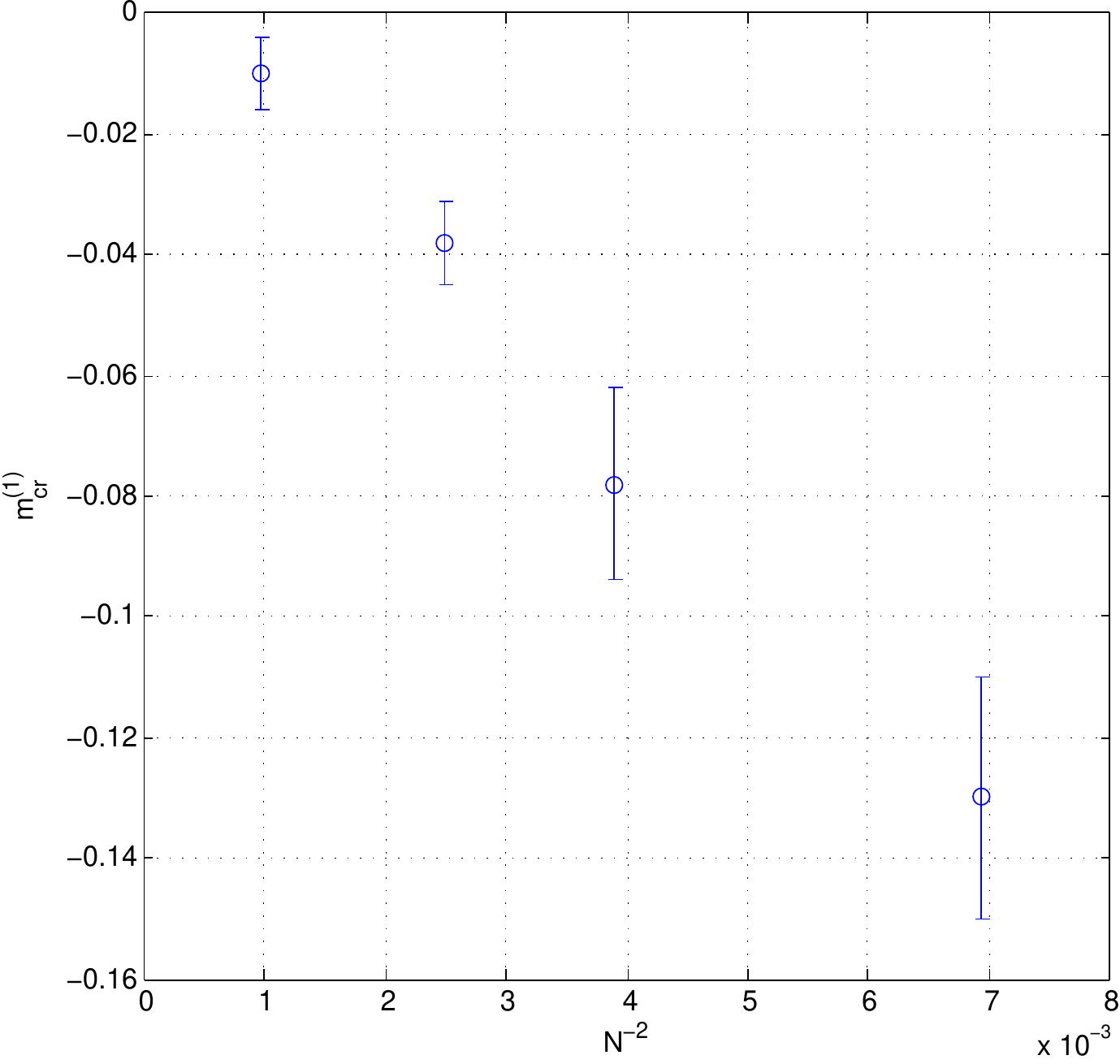}
     &
 \hspace{1.cm}
     \includegraphics[height=6.5cm,clip=true]{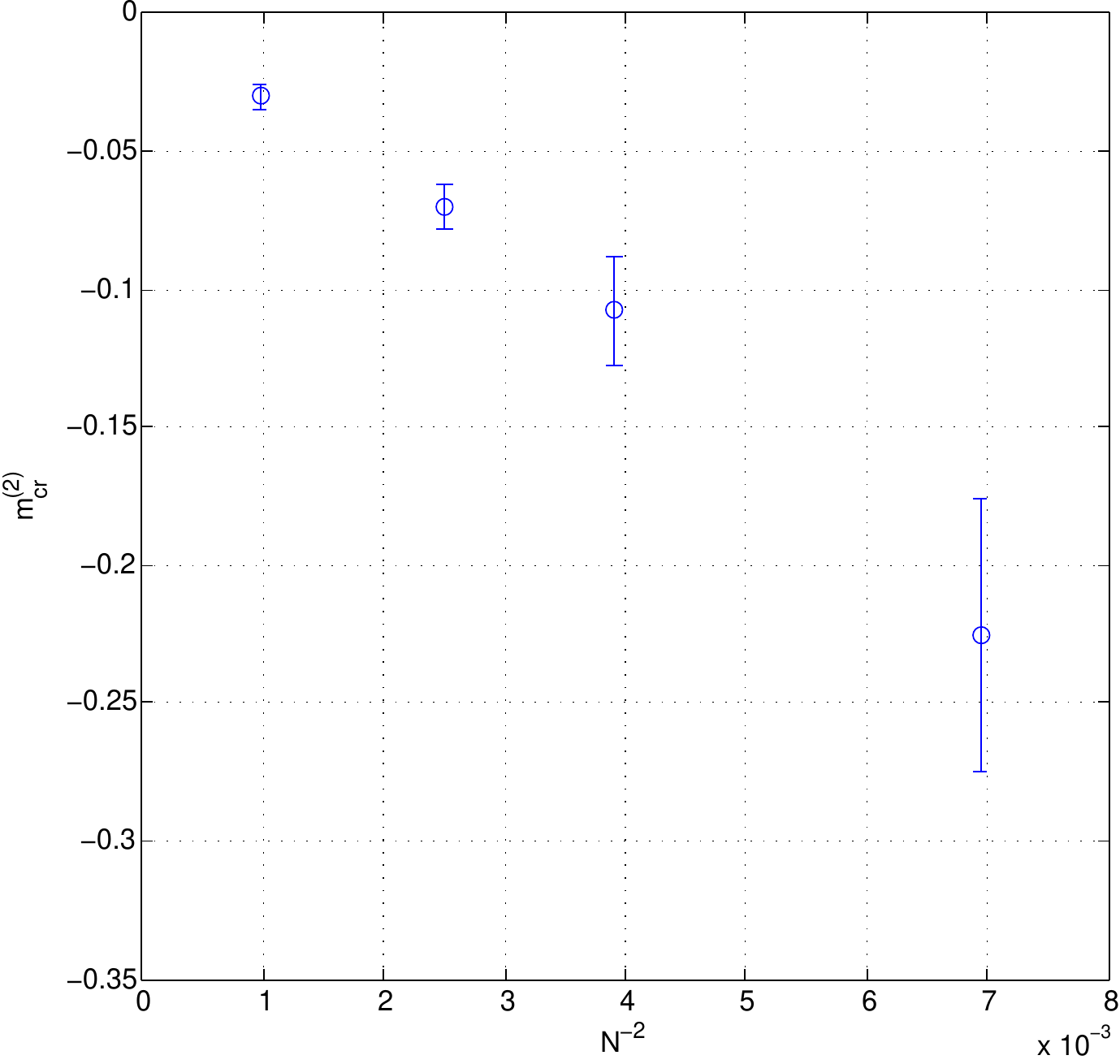}
  \end{tabular}
\end{center}
  \caption{Infinite volume extrapolations of the one-(left) and two-(right)loop
    critical mass; data are plotted (and extrapolated) as functions of
    an inverse power of the lattice size $N=L/a$.}
\end{figure}
 
Things are different at three-loop order. Since the critical mass
counterterm has been inserted up to two-loop order, 
$\hat{\Sigma_c}(\hat{p},\hat{m}_{cr},\beta^{-1})$ does not have to
extrapolate to zero. We get instead a first original result: as a 
byproduct of our computations, we can estimate the three-loop critical 
mass. Three-loop results are plotted in Figure~4. The infinite volume 
extrapolation of our results reads $\hat m_{cr}^{(3)} = -3.94 (4)$. 

\begin{figure}[!htb]
\begin{center}
  \begin{tabular}{cc}
     \includegraphics[height=6.5cm,clip=true]{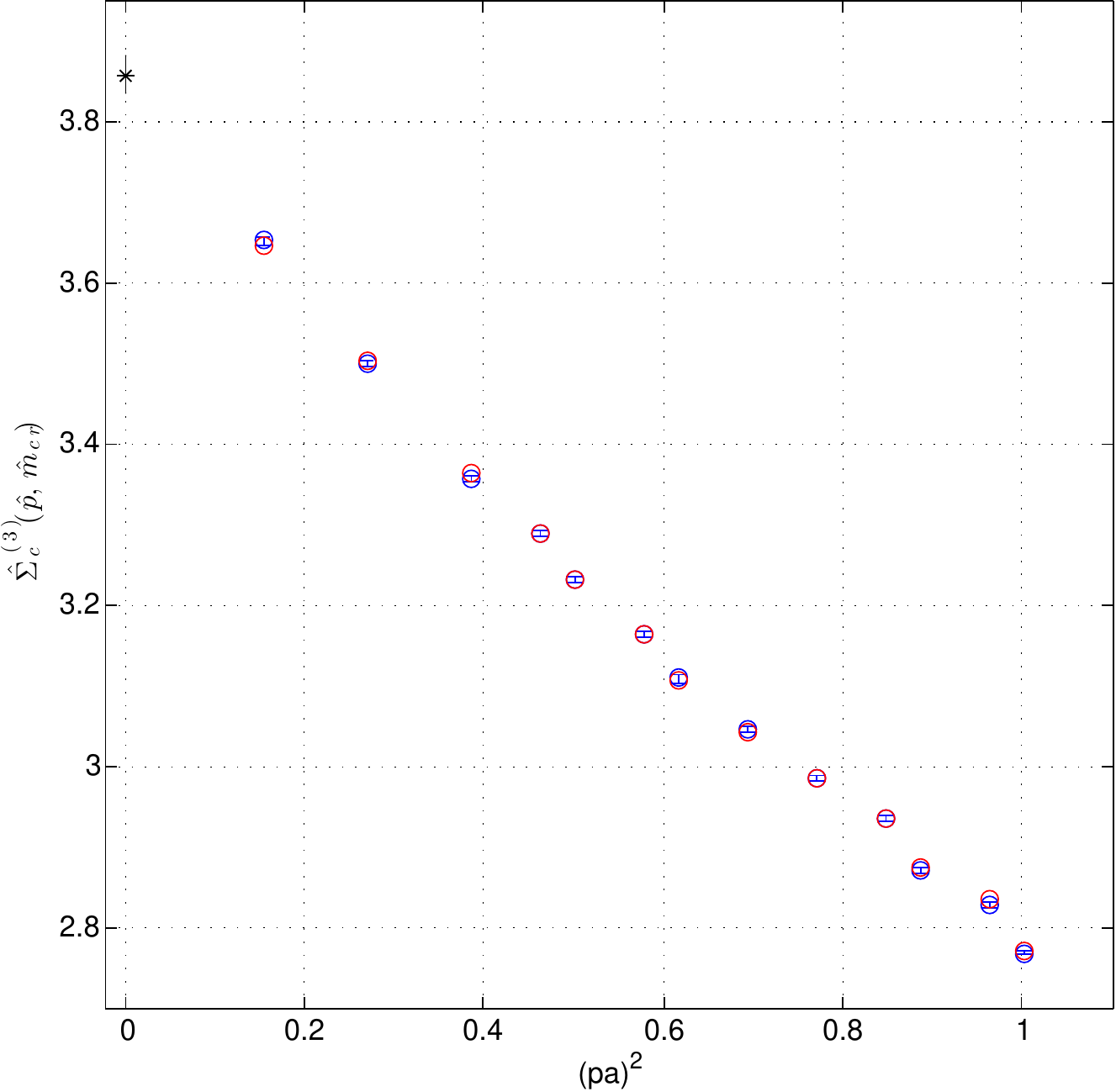}
     &
 \hspace{1.cm}
     \includegraphics[height=6.5cm,clip=true]{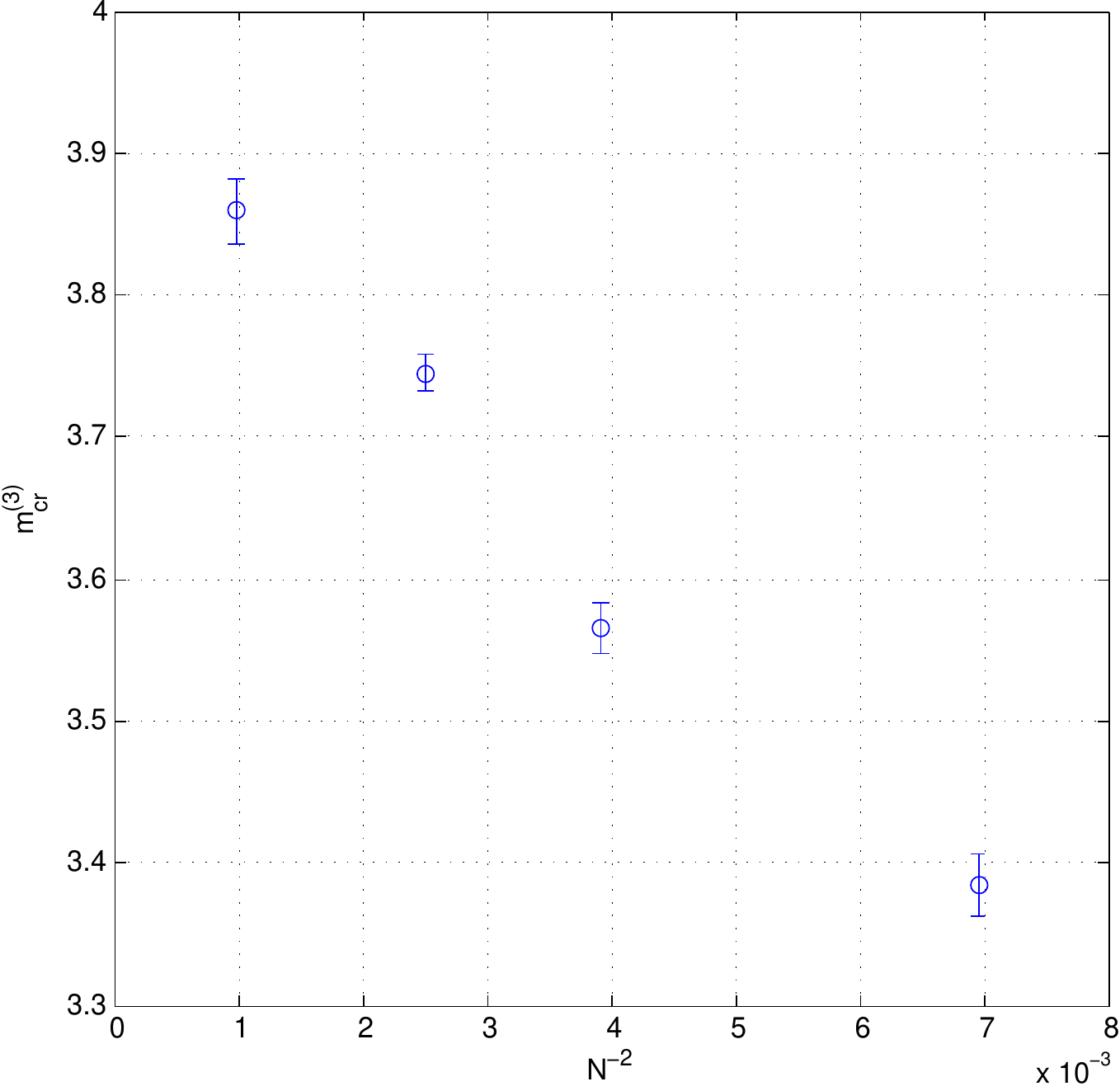}
  \end{tabular}
\end{center}
  \caption{Three-loop critical mass: zero momentum extrapolation 
on a $32^4$ lattice (left)
and infinite volume extrapolation (right).}
\end{figure}

\section{The continuum and infinite volume limits of the $Z_{O_{\Gamma}}$}

Our main goal is to compute the renormalization constants from our master
formula Eq.~(3). As said, in the $\beta^{-1}$ expansions of Eq.(16) 
the only unknown are the $z_n^{(0)}$: these are the quantities we are
interested in. 
The $O_\Gamma(p)$ of Eq.~(2) are the basic building blocks we have to 
compute; more precisely, we have to compute their lattice
counterparts. 
Their evaluation in NSPT is much the same as in the
non-perturbative case: computing them basically amounts to a fair
number of inversions of the Dirac operator (on convenient sources) 
and a fair number of 
scalar products. We point out that we always deal
with sources in momentum space. This is quite natural in our
computation environment, since the inversion of the Dirac operator
proceeds back and forth from momentum to configuration space
(see \cite{NSPT1}).

\subsection{Hypercubic symmetry and continuum limit: the case of $Z_q$}

\noindent To compute the various $Z_{O_\Gamma}$, the RI'-MOM master formula 
requires the knowledge of the field renormalization constant $Z_q$,
which is entailed in the self-energy via Eq.(1). 
The relevant component is the one along the gamma matrices, for which 
(in the infinite volume limit) 
hypercubic symmetry fixes the expected form 
\begin{equation}
\hat{\Sigma}_{\gamma} = \frac{1}{4} \sum_{\mu} \gamma_{\mu} {\rm Tr_{spin}}(\gamma_{\mu} \hat{\Sigma} ) 
= i\sum_\mu\gamma_\mu\hat p_\mu\left( 
\hat\Sigma_{\gamma}^{(0)}(\hat p)+ \hat p^2_\mu\hat\Sigma_{\gamma}^{(1)}(\hat p)+ \hat p^4_\mu\hat\Sigma_{\gamma}^{(2)}(\hat p) + \ldots \right)
\end{equation}
\noindent There is a tower of contributions on top of the one expected
in the continuum: they are due to the reduced symmetry and (as
expected) they are irrelevant ones, \ie they show up as power of $\hat
p_{\mu}^2 = (a p_{\mu})^2$. 
As another consequence of the lattice symmetry, each
$\hat\Sigma_{\gamma}^{(i)}(\hat p)$ is not only a function 
of $\sum_\nu \hat p_\nu^2$, but of all the possible 
hypercubic invariants, \eg those listed in Eq.~(20).\\

The prescription to get $Z_q$ at any scale $p$ 
is clear from the definition. Notice that, in the continuum limit, 
one can equivalently compute 
\begin{equation}
Z_q(\mu=p,\beta^{-1}) = -i\frac{1}{3}\frac{Tr(\gamma_{\bar \mu}S^{-1}(p))}{p_{\bar \mu}}.
\end{equation}
In the previous formula we have taken the shortcuts of recognizing $p$ as
the renormalization scale and of writing the dependence on the coupling
on the left-hand side only. 
Here $\bar \mu$ is any of the directions
(\ie $p_{\bar \mu}$ is any component of the momentum at hand).

Our computation proceeds by evaluating the right-hand side of
Eq.~(22), which in our setting does not equal the left-hand side, 
but yields (see Eq.~(21))
\begin{equation}
\widehat\Sigma_{\gamma}(\hat p, \bar \mu) \equiv \hat\Sigma_{\gamma}^{(0)}(\hat p)+ \hat p^2_{\bar \mu}\hat\Sigma_{\gamma}^{(1)}(\hat p)+ 
\hat p^4_{\bar \mu}\hat\Sigma_{\gamma}^{(2)}(\hat p) + \ldots 
\end{equation}
where the dependence on the choice of a given direction $\bar \mu$ 
is explicit. Notice that this is a sloppy notation, in which we are
assuming no finite size effects, which are certainly there in any
NSPT simulations: we will correct for them later. 
The result obtained at one-loop on a $32^4$ lattice can be 
inspected in Figure~5 (left). Data are plotted vs $\sum_\nu \hat p_\nu^2$: 
as we have already pointed out, they are not a function of this variable only, 
which is the reason why the curves are not completely smooth. 
Errors are negligible compared to the size of the symbols: this should
be born in mind in what follows. In order to connect what is plotted to 
a value for $Z_q$, a few general observations should be made:

\begin{itemize}
\item $\widehat\Sigma_{\gamma}(\hat p, \bar \mu)$ in general contains 
logarithms. Since we can not disentangle them from 
irrelevant contributions (this would require a terrific numerical precision),
we subtract them (they are known from the method discussed in
Section~2). This mechanism of subtracting the logs is a common feature
of our method and is in place for the computation of any
renormalization constant. Actually Figure~5 is with this respect 
a particular case, since
there is no log at one-loop order for the self-energy in Landau
gauge. In other words, if we plotted the two- or three-loop 
computations of the same 
quantity, then we should perform the subtraction. 
\item After (possibly) subtracting the logs, one is left with a variety of
  irrelevant contributions: only one number will survive in the
  continuum limit and one needs a procedure to extract it.
\item Irrelevant contributions are organized in such a way that {\em
    families} of curves are easily recognized: each different family
   is denoted by a different symbol in Figure~5. 
\end{itemize}

\begin{figure}[!tb]
\begin{center}
  \begin{tabular}{cc}
     \includegraphics[height=6.5cm,clip=true]{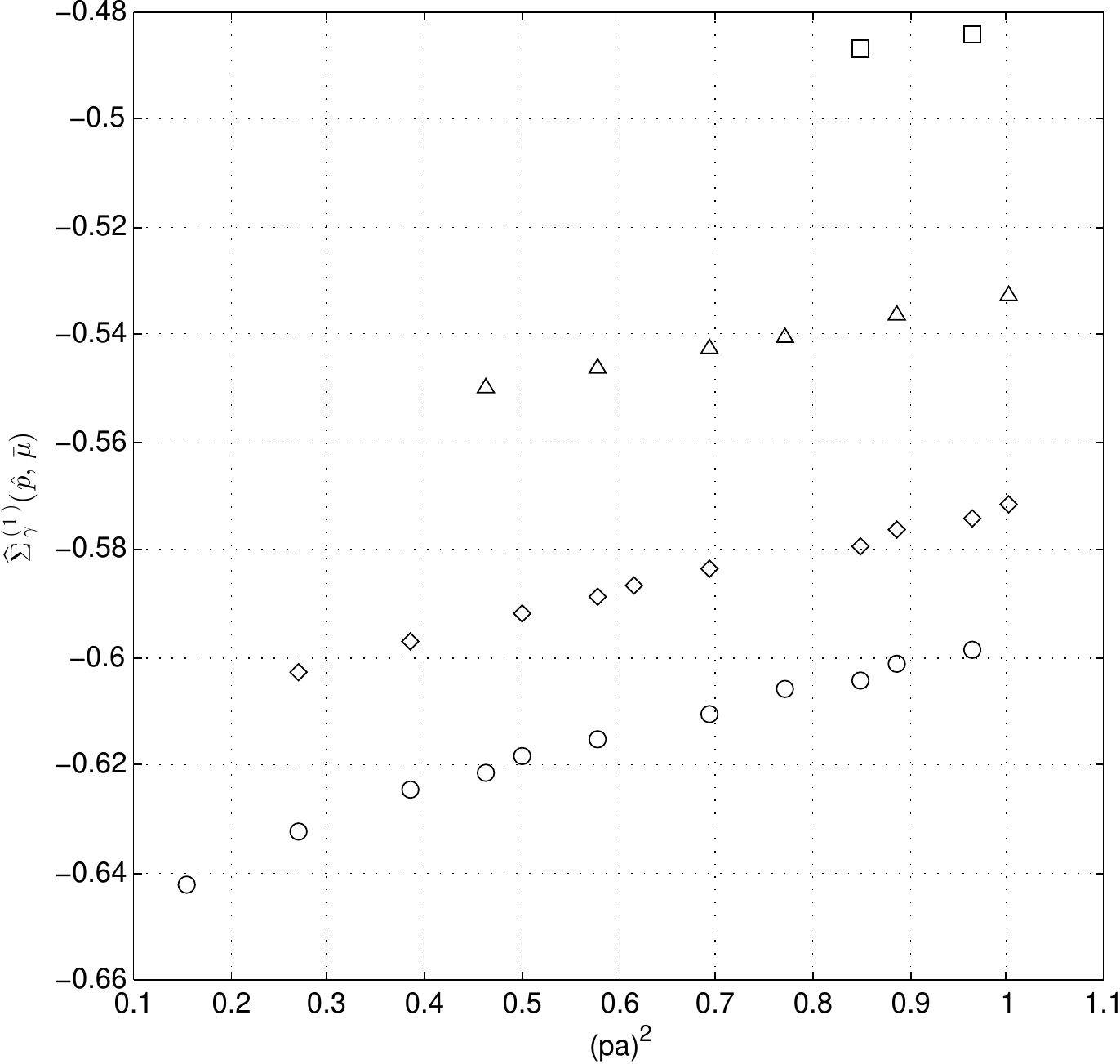}
     &
 \hspace{1.cm}
     \includegraphics[height=6.5cm,clip=true]{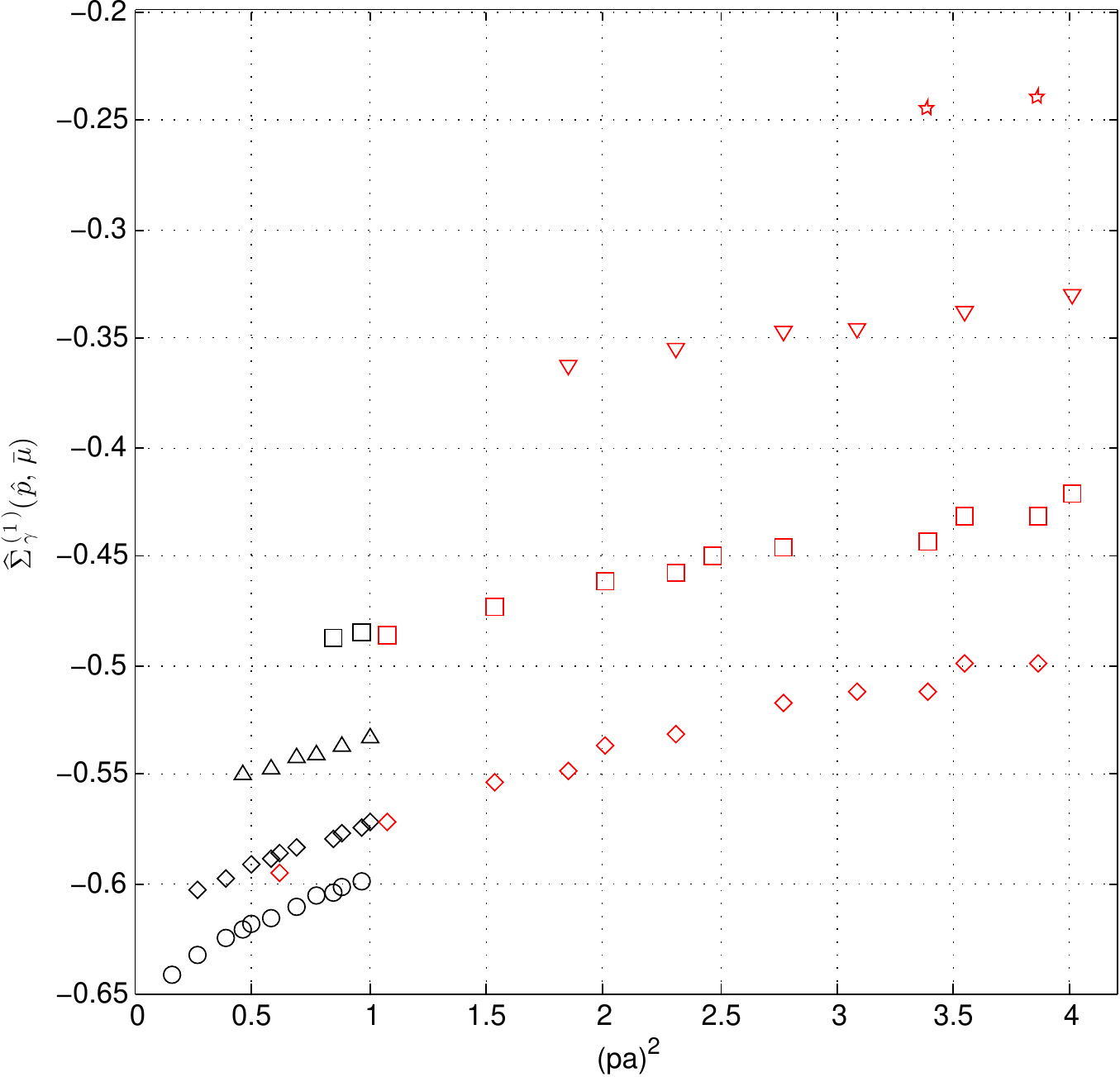}
  \end{tabular}
\end{center}
  \caption{One-loop evaluation of $\widehat\Sigma_{\gamma}(\hat p, \bar \mu)$
    (see Eq.~(23)) on a
    $32^4$ lattice (left) and on both a $32^4$ and a $16^4$ lattice (right).}
\end{figure}

\noindent 
Why data arrange in {\em families} is clear from Eq.~(23). 
On a finite lattice, the allowed (dimensionless) momenta are of the form
\beq
\hat p_{\mu} = \frac{2 \pi}{N} n_{\mu},
\eeq
where the $n_{\mu}$ are integer numbers.  
Given a 4-tuple $\{n_1,n_2,n_3,n_4\}$, the values of the scalar functions 
$\hat\Sigma_{\gamma}^{(i)}(\hat p)$ are fixed. Depending on the choice of the direction 
$\bar \mu$ in the right-hand side of Eq.~(22), one gets different 
combinations of the $\hat\Sigma_{\gamma}^{(i)}(\hat p)$ 
(depending on the length $|\hat p_{\bar \mu}|$) 
and thus different values for $\widehat\Sigma_{\gamma}(\hat p, \bar \mu)$. 
To be definite: suppose we pick the 4-tuple $\{1,1,1,2\}$.  
(this is the second lowest value of $\sum_\nu \hat p_\nu^2$ 
in Figure~5). When $\bar \mu = 1,2,3$ we 
get the point on the first family (by this we mean the lowest lying, \ie the 
circles), while for $\bar \mu = 4$ we get the point on the second family (\ie 
the diamonds). In other words, there is a different family for each
different value of the length $|\hat p_{\mu}|$.\\

As a first attempt, we can fit Eq.~(23) to our data by first
(possibly) subtracting leading logs and then 
taking for each 
$\hat\Sigma_{\gamma}^{(i)}(\hat p) |_{\mbox{\tiny{log subtr}}}$ 
an expansion in hypercubic invariants, \eg
\beq
\hat\Sigma_{\gamma}^{(i)}(\hat p) |_{\mbox{\tiny{log subtr}}} = 
c_1^{(i)} +
c_2^{(i)} \sum_\nu \hat p_\nu^2 +
c_3^{(i)} \frac{\sum_\nu \hat p_\nu^4}{\sum_\nu \hat p_\nu^2} + 
\mathcal{O}(a^4).
\eeq

\noindent A trivial power-counting fixes the order at which each 
$\hat\Sigma_{\gamma}^{(i)}(\hat p)$ is expanded (there is a factor of $\hat 
p^{2i}_{\bar \mu}$ in front). The only term surviving the $a\to 0$ limit is $c_1^{(0)}$, 
which is the only one we are interested in. Referring to the data of
Figure~5, it is the estimate of $z_{q1}^{(0)}$, the finite part of the one-loop 
$Z_q$, as obtained from the computation on a $32^4$ lattice. \\

We do not report results for $Z_q$: in the following the dependence on
$Z_q$ of Eq.~(3) will be eliminated by making use of the quantity 
$\widehat\Sigma_{\gamma}(\hat p, \bar \mu)$ defined in Eq.~(23), but before
we can deal with this, we have to address the finite volume effects issue.

\subsection{General procedure: disentangling finite volume effects}

In Figure~5 (right) we display the computation of the one-loop 
$\widehat\Sigma_{\gamma}(\hat p, \bar \mu)$ 
on both a $32^4$ (black symbols) and a $16^4$ (red symbols) lattice. 
The 4-tuples $n_{\mu}$ defining the momenta are the same for 
measurements on both lattice sizes (as one can see, there is the 
same number of black and red symbols). 
{\em Families} join quite smoothly, in a way which is dictated by Eq.~(24). 
Suppose we pick the same 4-tuple $\{1,1,1,2\}$\footnote{We recall that this is 
the second lowest value of $\sum_\nu \hat p_\nu^2$ 
for both lattice sizes in Figure~5.} 
both on $32^4$ and on $16^4$ and inspect the values of 
$\widehat\Sigma_{\gamma}(\hat p, \bar \mu)$: we have to look for 
them {\em (a)} at different values of the abscissa $\sum_\nu \hat p_\nu^2$ 
and {\em (b)} on different families. To make the last point even
clearer, 
consider the choice $\bar \mu = 1$: on $32^4$ it results in 
$|\hat p_{\bar \mu}| = \pi/16$ and makes $\widehat\Sigma_{\gamma}(\hat p, \bar\mu=1)$ 
fall on the first family (circles);  $\bar \mu = 1$ on $16^4$ results in 
$|\hat p_{\bar \mu}| = \pi/8$ and makes $\widehat\Sigma_{\gamma}(\hat p, \bar\mu=1)$ 
fall on the second family (diamonds). \\

If there were no finite size effects, all the families should join in
a perfectly smooth way and there should be a few points falling 
exactly on top of each other, \eg the value of 
$\widehat\Sigma_{\gamma}(\hat p, \bar\mu)$ for the 4-tuple $\{1,1,1,1\}$ 
on $16^4$ should fall on top of the one associated to the 
4-tuple $\{2,2,2,2\}$ on $32^4$. By inspecting the data 
($\{1,1,1,1\}$ corresponds to the lowest value of $\sum_\nu \hat p_\nu^2$ 
for both lattice sizes) we see that the red diamond does not fall exactly on top of 
black one. This is a first hint at some finite size effect. \\

To correct for finite size effects we proceed along the lines that
were first introduced in \cite{NSPT_Gh,NSPT_Gl}. We 
infer a $pL$ dependence in 
$\widehat\Sigma_{\gamma}$ (this is expected on dimensional grounds) and
define a correction with respect to the infinite volume result by
simply adding and subtracting the latter
\begin{eqnarray}
\widehat\Sigma_{\gamma}(\hat p, pL,\bar \mu) & = & \widehat\Sigma_{\gamma}(\hat p,
\infty,\bar \mu) + \left( \widehat\Sigma_{\gamma}(\hat p, pL,\bar \mu) -
  \widehat\Sigma_{\gamma}(\hat p, \infty,\bar \mu) \right) \nonumber \\
& \equiv & \widehat\Sigma_{\gamma}(\hat p, \infty,\bar \mu) + \Delta\widehat\Sigma_{\gamma}(\hat p, pL,\bar \mu)
\end{eqnarray}
To a first approximation we now let
\beq
\Delta\widehat\Sigma_{\gamma}(\hat p, pL,\bar \mu) \sim  \Delta\widehat\Sigma_{\gamma}(pL),
\eeq
the main rationale being that we neglect {\em corrections on top of
  corrections} (more on this later). As a result, we have a better form to be fitted to 
$\widehat\Sigma_{\gamma} |_{\mbox{\tiny{log subtr}}}$, \eg (we here
assume a low order 
expansions in terms of $a$, actually lower than the ones we typically manage)
\beq
\widehat\Sigma_{\gamma}(\hat p, pL,\bar \mu)  |_{\mbox{\tiny{log subtr}}} = 
c_1^{(0)} +
c_2^{(0)} \sum_\nu \hat p_\nu^2 +
c_3^{(0)} \frac{\sum_\nu \hat p_\nu^4}{\sum_\nu \hat p_\nu^2} +
c_1^{(1)} p^2_{\bar \mu} + 
\Delta\widehat\Sigma_{\gamma}(pL) + \mathcal{O}(a^4).
\eeq
The above formula opens the way to a combined fit of measurements on
different lattice sizes: since 
\[
p_\mu L = \frac{2\pi n_\mu}{L}L = 2\pi n_\mu,
\]
there is only one finite size correction for each 4-tuple
$n_\mu$ and no functional form has to be inferred for the
correction. 

The quantities $\widehat\Sigma_{\gamma}(\hat
p, pL, \bar\mu)$ have been taken as examples to clarify the way we deal with the
extraction of the continuum and infinite volume limits, but they
are not directly used to determine the field
renormalization $Z_q$. Without making explicit reference to $Z_q$ 
we instead rewrite the 
{\em finite part} of the currents RI'-MOM renormalization 
constants as 
\beq
Z_{O_\Gamma}(\mu=p,\beta^{-1}) |_{\mbox{\tiny{finite part}}}  = 
\lim_{\substack{a \rightarrow 0 \\ L \rightarrow \infty}} 
\frac{\widehat\Sigma_{\gamma}(\hat p, pL, \bar \mu)}
{\hat O_{\Gamma}(\hat p, pL)} |_{\mbox{\tiny{log subtr}}}
\eeq
The right-hand side has to be evaluated order by order. 
The dependence on $Z_q$ has been traded for the 
$\widehat\Sigma_{\gamma}(\hat p, pL, \bar \mu)$, which reconstructs 
the $Z_q$ contribution to the left-hand side 
in the limits which are taken in Eq.~(29). Before taking
the limits this provides
a lot of irrelevant contributions (and finite size effects) on top of
what is {\em per se} contained in the $\hat O_{\Gamma}(\hat p, pL)$, which
are the finite lattice version of the  $O_{\Gamma}(p)$ of
Eq.~(3). Here there is a subtlety connected with the dependence on 
direction of vector and axial currents: we will comment on this. 
Notice finally the
notation $\ldots |_{\mbox{\tiny{log subtr}}}$: 
this means that the leading logarithms which plagues 
$Z_{O_\Gamma}(\mu=p,\beta^{-1})$ as a function of $pa$ have been
subtracted (once again, they are known from the prescriptions of
Section~2). The 
$a \rightarrow 0$ limit of Eq.~(29) can be taken only provided 
this subtraction is performed. \\

Let's consider the vertex function relevant for the computation of the
vector current. In the continuum one has
\[
\Gamma_V^{\mu}(p) = \gamma^{\mu} \, \Sigma_V^{(1)}(p^2) +
\frac{p^{\mu}\slashed{p}}{p^2} \, \Sigma_V^{(2)}(p^2)
\]
where the extra contribution (with respect to the tree level
structure) vanishes at $p=0$. 
The lattice version, due to the same 
mechanism which is place for 
$\hat\Sigma_{\gamma}$ (\ie reduced symmetry
with respect to the continuum), reads
\[
\hat\Gamma_V^{\mu}(\hat{p}) = \gamma^{\mu} \, \hat\Sigma_V^{(1)}(\hat{p}) +
\frac{\hat{p}^{\mu}\sum_{\nu}\gamma_{\nu}\hat{p}_{\nu}}{\sum_{\nu} \hat{p}_{\nu}^2} \, \Sigma_V^{(2)}(\hat{p})+
\frac{\hat{p}^{\mu 3}\sum_{\nu}\gamma_{\nu}\hat{p}_{\nu}}{\sum_{\nu} \hat{p}_{\nu}^2} \, \Sigma_V^{(3)}(\hat{p})+
\frac{\hat{p}^{\mu}\sum_{\nu}\gamma_{\nu}\hat{p}_{\nu}^3}{\sum_{\nu} \hat{p}_{\nu}^2} \, \Sigma_V^{(4)}(\hat{p})+
\ldots
\]
where we have only written the first two irrelevant
extra-contributions; the 
$\Sigma_V^{(i)}(\hat{p})$ depend on all the hypercubic invariants. If
we now choose $\gamma_{\mu}$ as the projector $\hat P_{O_{\Gamma}}$ 
of Eq.~(2) we would get a dependence on the direction 
(\ie we would get a $\hat O_{\Gamma}(\hat p, pL,\bar \mu)$, in the
notation which should by now be familiar). While this is not {\em per
  se} a conceptual problem, it is a practical one, since we would
have a very large set of coefficients to fit. We can consider
\[
\frac{1}{12} \left( \Tr\gamma_{\mu} \hat\Gamma_V^{\mu}(\hat{p}) - 
\Tr \frac{\hat{p}^{\mu}\sum_{\nu}\gamma_{\nu}\hat{p}_{\nu}}{\sum_{\nu}
  \hat{p}_{\nu}^2} \hat\Gamma_V^{\mu}(\hat{p}) \right)
\;\;\;\;\;\;\mbox{or} \;\;\;\;\;\; \frac{1}{4} \sum_{\mu} \Tr\gamma_{\mu}
  \hat\Gamma_V^{\mu}(\hat{p})
\]
in order to have no dependence on direction. In the first case the
contribution from $\Sigma_V^{(2)}(\hat{p})$ is eliminated at any value
of the momentum, but we verified that the two
procedures returns consistent results once irrelevant contributions
are extrapolated away. \\
The procedure of averaging over directions would cancel the dependence
on direction also in $\widehat\Sigma_{\gamma}(\hat p, pL, \bar \mu)$,
but as we saw this dependence provides a very effective handle for
assessing finite size effects. All in all, by retaining dependence on
direction in $\widehat\Sigma_{\gamma}(\hat p, pL, \bar \mu)$ and
eliminating a possible dependence on direction in 
$\hat O_{\Gamma}(\hat p, pL)$ we aim at a fit which is both effective
and not too demanding in terms of number of parameters.

From our NSPT computations 
we can evaluate (order by order) the quantities
\beq
\widehat O_\Gamma(\hat p, pL, \bar \mu)  \equiv 
\frac{\widehat\Sigma_{\gamma}(\hat p, pL, \bar \mu)}
{\hat O_{\Gamma}(\hat p, pL)} |_{\mbox{\tiny{log subtr}}}
\eeq
where (compares to Eq.~(29)) no limit is 
taken. The continuum and infinite volume limits are reconstructed by
first computing the $\widehat O_\Gamma(\hat p, pL, \bar \mu)$ at different 
values of $(\hat p, pL, \bar \mu)$ and then fitting the results
accordingly to the procedure we described above, for which we now
provide a few extra details (see also \cite{NSPT_Gh,NSPT_Gl}). 
In practice we proceed as follows:
\begin{itemize}
\item A given $\widehat O_\Gamma(\hat p, pL, \bar \mu)$ is computed on
  different sizes $N=L/a$ and for different values of lattice
  momenta. We stress once again that the definition of 
$\widehat O_\Gamma(\hat p, pL, \bar \mu)$ entails the subtraction of
leading logs contributions. 
\item We select a given interval of values of $\sum_\nu \hat p_\nu^2$: 
let's call it $I_{p2}$. We also decide 
a given order for our fit in terms of power of $pa$. 
This fixes the number of parameters we have to determine as for irrelevant
$pa$ contributions (we recall once again that this is fixed by
hypercubic symmetry). 
\item Given the momentum interval and the sizes $N$, we define 
  the set of points (\ie 4-momenta) to be fitted 
by first selecting 4-tuples $\{n_1,n_2,n_3,n_4\}$ 
satisfying the following requirement: the corresponding 4-momenta
should return a value of $\sum_\nu \hat p_\nu^2$ within the 
selected momentum interval $I_{p2}$ for each value of $N$. Notice that 
within the approximation of Eq.~(27) 
there is one parameter to be fitted for each of these 4-tuples: 
let's call it $\Delta \widehat O_\Gamma(pL)$. 
\item We add to the set of momenta selected above the 
measurement taken on the largest lattice size $N_{\mbox{\tiny{max}}}$ 
at the 4-momentum corresponding to the largest value of 
$\sum_\nu \hat p_\nu^2$ within $I_{p2}$. This data point is assumed
free of finite size effects and acts as a normalization point for our
fitting procedure (we study the stability of the fit by allowing the
inclusion of two data points from the largest lattice size, \ie 
the two 4-momenta corresponding to the largest and second 
largest value of $\sum_\nu \hat p_\nu^2$ within $I_{p2}$). 
\item The functional form of our fit (and the number of parameters to
  be determined) is now completely fixed. In particular, no attempt is
  made to fit {\em subleading logs}. 
\end{itemize}
One has to be aware that a number of assumptions were made, so
that the effectiveness of the fit has always to be assessed a
posteriori (at one-loop we can compare to results available from standard
techniques; at higher loops we can assess stability of the procedure 
and inspect the values of standard
indicators, \eg values of $\chi^2$). 

We plot in Figure~6 the one-loop $\widehat O_S(\hat p, pL, \bar\mu)$ 
(the $\widehat O_\Gamma(\hat p, pL, \bar \mu)$
for the scalar current). On the left, data 
are plotted as obtained on $32^4$ (black symbols) and $16^4$ (red
symbols), without any finite size corrections. Finite size effects are
manifest: red and black diamonds fail to fall on a smooth curve, and
the same holds for black and red squares 
(these are supposed to be {\em families} in
the jargon we introduced). On the right, we display the same
data corrected for finite size effects: the corrections have been
fitted according to our simplest recipe (in the spirit of
Eq.~(27)). Black and red diamonds and black and red squares now do
fall on smooth curves. The effectiveness of the fit is displayed in
Figure~7, where in particular one can see how well we determine
the final result we are interested in, \ie $z_{S1}^{(0)}$, in the notation
of Eq.~(16) (we recall that this is the counterpart of $c_1^{(0)}$ 
of Eq.~(28)).

\begin{figure}[!b]
\begin{center}
  \begin{tabular}{cc}
     \includegraphics[height=6.5cm,clip=true]{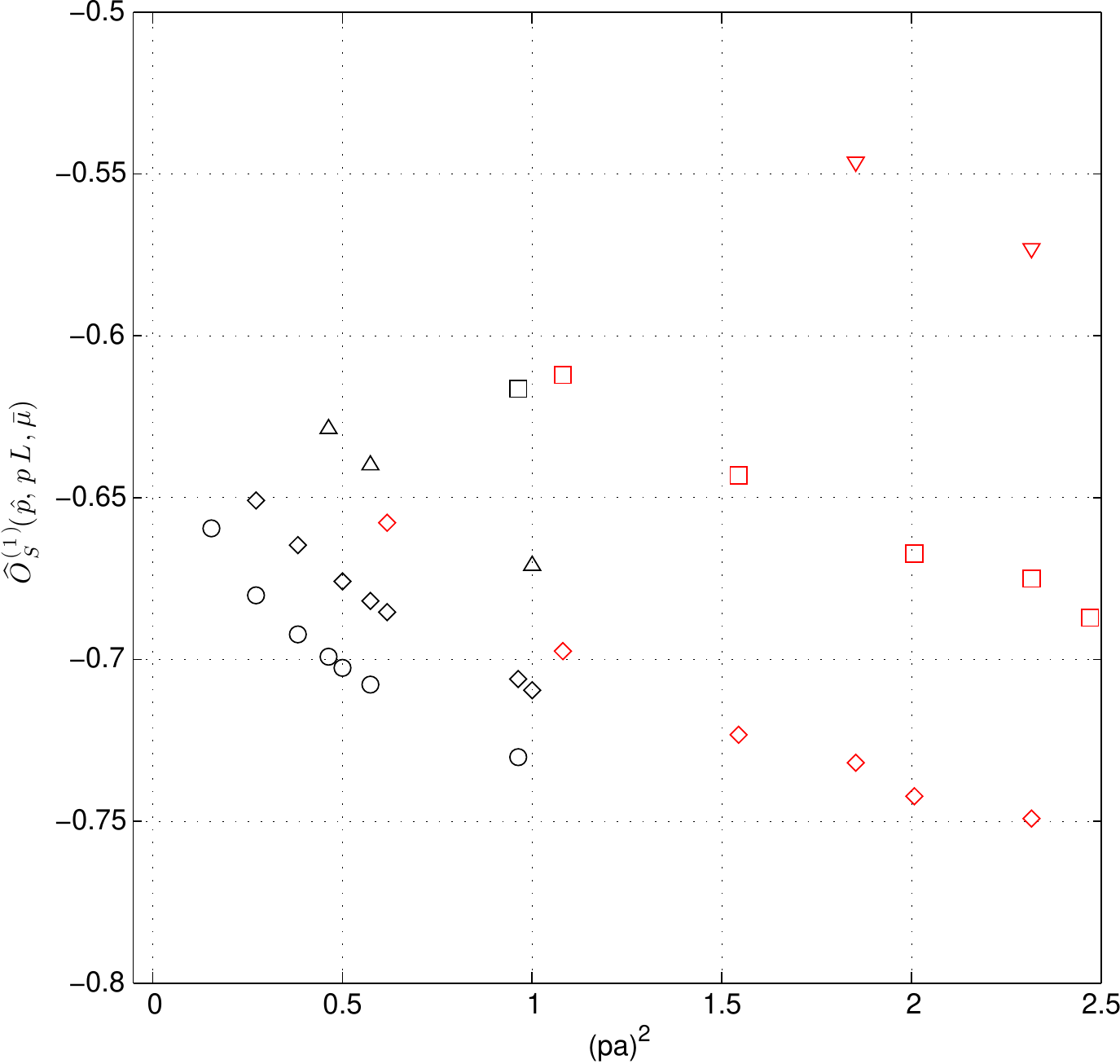}
     &
 \hspace{1.cm}
     \includegraphics[height=6.5cm,clip=true]{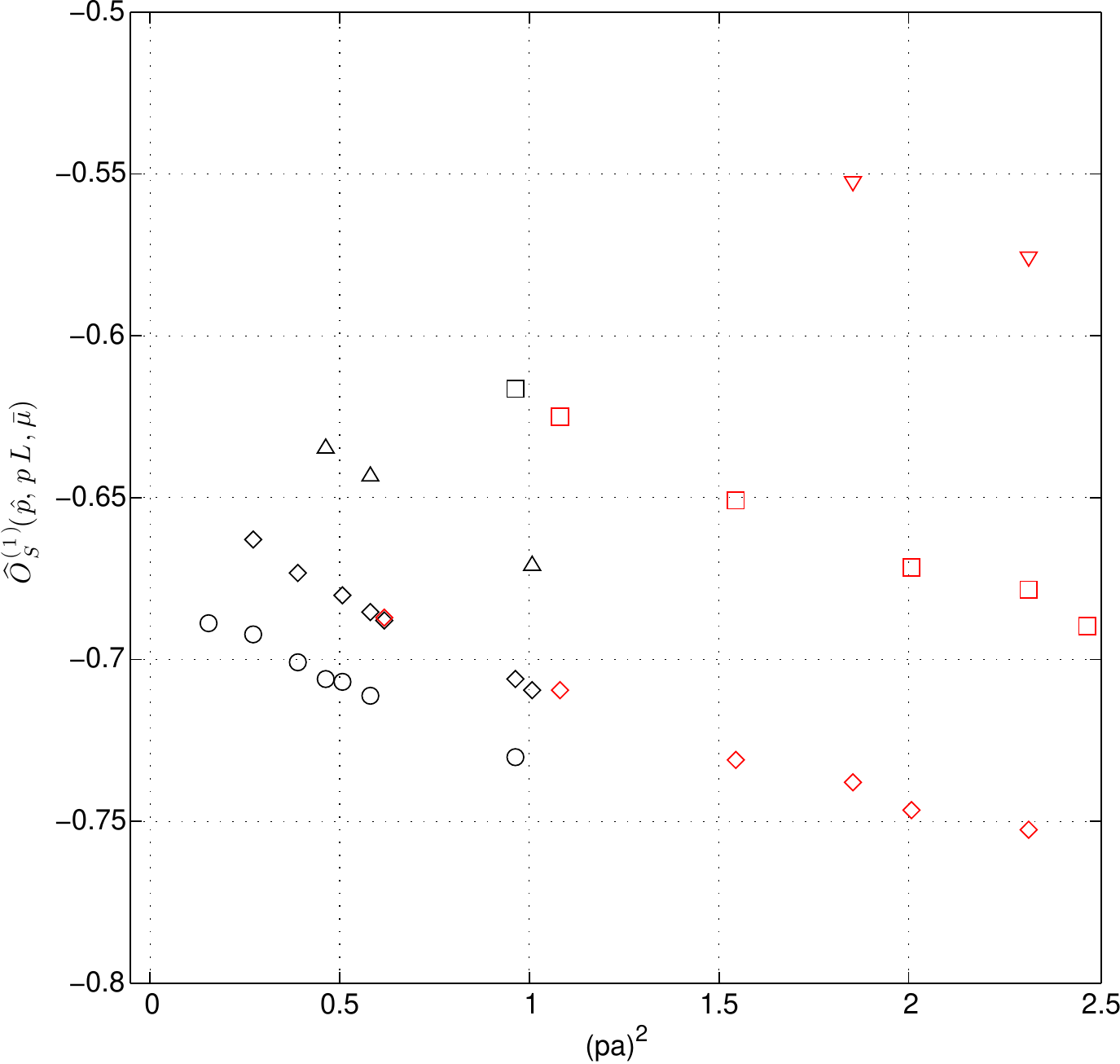}
  \end{tabular}
\end{center}
  \caption{One-loop $\widehat O_S(\hat p, pL, \bar\mu)$ (see Eq.~(30))
    measured on a $32^4$
    (black) and a $16^4$ (red) lattice,  
    without (left) and with (right) finite size corrections.}
\end{figure}


 Notice that finite size effects are more manifest in Figure~6 than in 
Figure~5. The latter refers to a quantity for which there is no log
involved (the one-loop field anomalous dimension vanishes in 
Landau gauge). In \cite{NSPT_Zs} 
it was observed that whenever logs are in place, finite size effects
can be quite large. This is a rationale in support of the 
strong assumption contained in Eq.~(27), which states that
we look for one single parameter ($\Delta \widehat O_\Gamma(pL)$) to 
correct for finite size effects in $\widehat O_\Gamma(\hat p, pL, \bar
\mu)$. In the definition of the latter a subtraction of leading logs
is in place. One can infer that on a finite lattice logarithmic
divergences are actually regulated in the IR (see the discussion of 
\cite{NSPT_Zs} in terms of {\em tamed logs}). As a matter of fact, the
finite size corrections that we get for finite renormalization
constants (\ie those of the vector and axial currents) are small, and
results are within errors quite consistent with those obtained by taking
into account only $32^4$ data.

\begin{figure}[!t]
\begin{center}
     \includegraphics[height=6.5cm,clip=true]{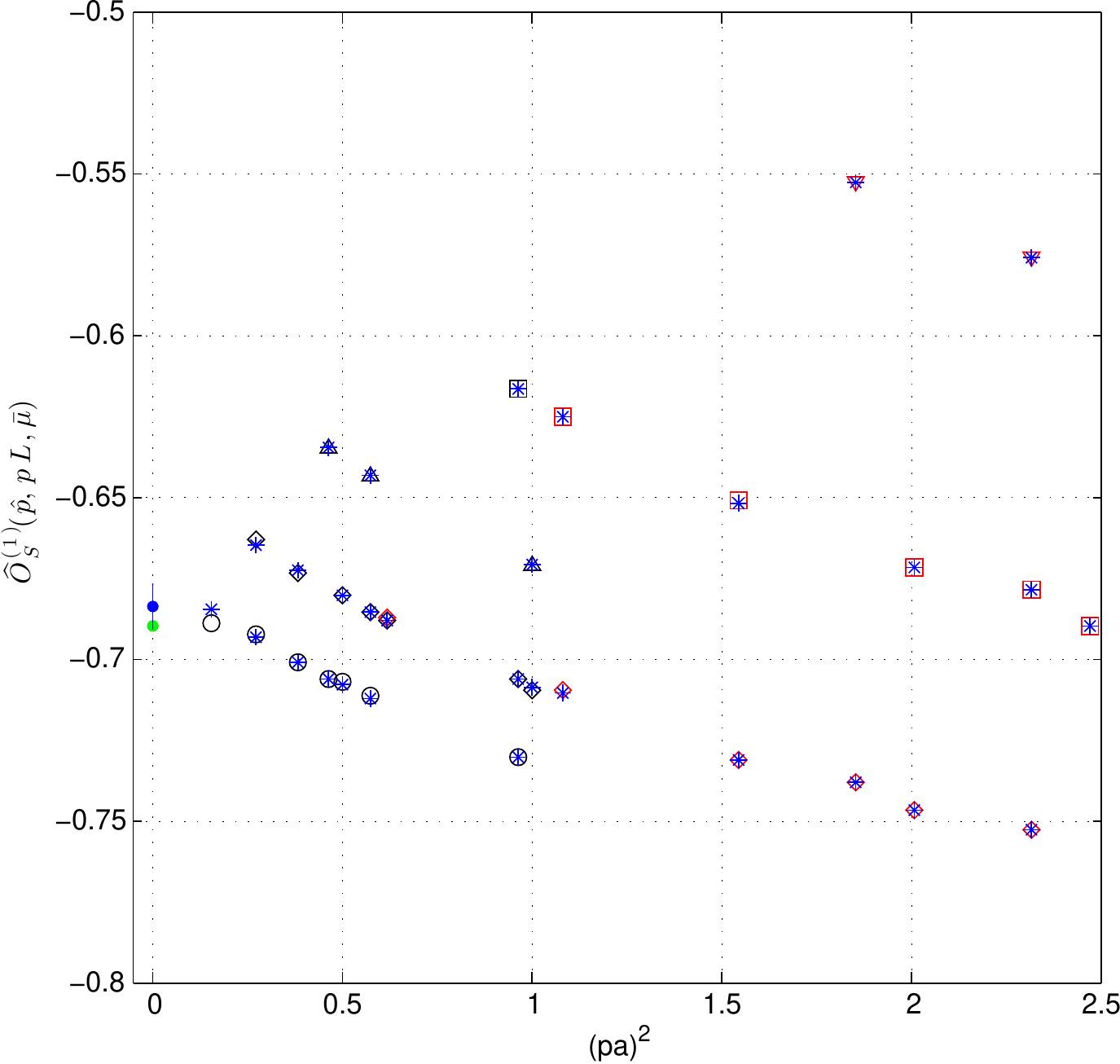}
\end{center}
  \caption{One-loop $\widehat O_S(\hat p, pL, \bar\mu)$ (see Eq.~(30))
    measured on a $32^4$
    (black) and a $16^4$ (red) lattice, with finite size corrections 
    and the fitted form plotted on top of data}
\end{figure}

\section{Results: three-loop expansion of $Z_S$, $Z_P$, $Z_V$, $Z_A$}

In Table~2 we report the coefficients of the three-loop expansion of 
$Z_S$, $Z_P$, $Z_V$ and $Z_A$\footnote{Comparing to the preliminary 
results in \cite{NSPT_Zs_LAT2012} the reader will recognize a typo for 
$Z_P$ at second loop.}. 
The expansion parameter is $\beta^{-1}$.
We quote the analytical one-loop results \cite{Aoki}: the 
comparison confirms the effectiveness of our method. 
Errors are dominated by the stability of fits with respect to the
change of fitting ranges, functional forms, number of lattice sizes
simultaneously taken into account. Three-loop results
for $Z_S$ and $Z_P$ have the indetermination in the coupling matching 
as an extra source of error (see discussion in Section~3).

Non-perturbative results for the quantities we computed are available
in \cite{ETMC_Zs}, where Twisted-Mass fermions regularization is in
place for the same (RI'-MOM) renormalization scheme. The latter is
defined in the chiral limit, and hence results must be the same as for
Wilson fermions (our case). In order to make a comparison, we have to
sum our series. The computations of \cite{ETMC_Zs}  are at three 
values of $\beta$, namely $3.8$, $3.9$, $4.05$. The last one is
in principle the best suited for a comparison in PT.

\begin{table}[!t]
\caption{One-, two- and three-loop coefficients of the renormalization 
constants for quark bilinears. Expansions are in
$\beta^{-1}$. One-loop analytical results are reported for comparison.}
\begin{center}
\begin{tabular}{|c|c|c|c|c|}
\hline
\, & \small {\em analytical} & \, & \, & \,\\ 
\,  & \small {\em one-loop} & one-loop & two-loop & three-loop \\
\hline
\hline
$Z_S$ & -0.6893 & -0.683(7) & -0.777(24) & -1.96(14) \\
$Z_P$ & -1.1010 & -1.098(11) & -1.299(38) & -3.19(21) \\
$Z_V$ & -0.8411 & -0.838(6) & -0.891(17) & -1.870(65) \\
$Z_A$ & -0.6352 & -0.633(4) & -0.611(16) & -1.198(57) \\
\hline
\hline
\end{tabular}\\
\end{center}
\end{table}

The simplest (and straightforward) recipe is to sum the series in the
coupling $\beta^{-1}$. At $\beta=4.05$ this results in 
$Z_V(\beta=4.05) = 0.710(28)$, $Z_A(\beta=4.05) = 0.788(18)$, 
$Z_S(\beta=4.05) = 0.753(30)$, $Z_P(\beta=4.05) = 0.610(48)$.
Here we adhere to a standar recipe for pinning down an (order of
magnitude for the) error: it is the three-loop contribution 
itself\footnote{Notice that this largely dominates the error 
which is coming from the errors on the different coefficients
themselves.}. At this stage one can inspect not too big discrepancies
with the values reported in \cite{ETMC_Zs} for the finite
renormalization constants (actually $Z_A$ is {\em de facto} fully
consistent). Larger deviations are inspected for $Z_S$ and $Z_P$.

It has become very popular \cite{LepMack} to sum LPT results 
making use of different couplings, a procedure that is often 
generically referred to as {\em Boosted Perturbation Theory} (BPT). 
As our group also observed in \cite{NSPT_Zs}, there is a large 
fraction of arbitrariness in one-loop BPT. Having three-loop 
results available provides a by far more stringent
control. We will devote much more space to this in
\cite{NSPT_Iwa}. Here we merely quote that the use of different
couplings (much the same was done in \cite{NSPT_Zs}) makes 
$Z_V$ and $Z_A$ fully consistent with the non-perturbative 
results of \cite{ETMC_Zs}, while for $Z_S$ and $Z_P$ the 
agreement improves, but it is not full. We notice that these are the
cases for which in our case it was crucial to look for finite size
effects corrections.

\section{Conclusions and prospects}

The main message of this paper is that computing three-loop
renormalization constants for Lattice QCD is fully viable, both for
finite and for logarithmically divergent quantities. The control on
both finite lattice spacing and finite volume effects appears 
solid and reliable. 

While three-loop finite renormalization constants are fully consistent
with non-perturbative results, the logarithmically divergent $Z_S$ and
$Z_P$ are not. One should keep in mind that a typical RI'-MOM lattice 
computation is performed over a range of momenta eventually pinching
the IR region. Now, contributions 
at low momenta are from one side supposed to be
relevant in assessing irrelevant (UV) contributions (higher powers
of $pa$ are suppressed), from another side 
prone to suffer from finite size (IR) effects. All in all, there is a subtle
interplay of UV and IR effects. 

We proposed a computational scheme to take control over this 
interplay in our perturbative framework. In \cite{NSPT_Iwa}, which is
dedicated to three-loop computations in a different gluonic
regularization (Iwasaki action), we devote more attention to quantify
the impact of irrelevant and finite volume contributions, comparing
results for the two different gluonic action (TLS and Iwasaki). It is
important to keep in mind that the numerics of IR effects that we 
compute in LPT are not supposed to be the same of non-perturbative
computations, while there is quite a consensus that this is the case
for irrelevant (UV) ones. In \cite{NSPT_Iwa} we discuss what of our
approach could be relevant for the non-perturbative framework in terms
of methodology.

\section*{Acknowledgments}
\par\noindent
We warmly thank Luigi Scorzato for the long-lasting collaboration on
NSPT and Christian Torrero, who has taken part in the long-lasting
project of three-loop computation of LQCD renormalization constants. 
We are very grateful to M. Bonini, V. Lubicz, C. Tarantino, R. Frezzotti, P. Dimopoulos 
and H. Panagopoulos for stimulating discussions. \\
This research is supported by the Research Executive Agency (REA) of the European Union under
Grant Agreement No. PITN-GA-2009-238353 (ITN STRONGnet).
We acknowledge partial support from both Italian MURST 
under contract PRIN2009 (20093BMNPR 004) and from I.N.F.N. under {\sl
  i.s. MI11} (now {\em QCDLAT}). 
We are grateful to the AuroraScience Collaboration for the computer time that was made available on the
Aurora system. 
It is a pleasure for F.D.R. to thank the Aspen Center for Physics for hospitality 
during the 2010 summer program: the long process of preparation of this work had a 
substantial progress on those days.


\end{document}